\documentclass[english,aps,twocolumn,showpacs,prb]{revtex4}
\usepackage[T1]{fontenc}
\usepackage[latin9]{inputenc}
\usepackage{textcomp}
\usepackage{amssymb}
\usepackage{graphicx}
\usepackage{esint}

\makeatletter

\newcommand{\lyxdot}{.}

\@ifundefined{textcolor}{}
{%
 \definecolor{BLACK}{gray}{0}
 \definecolor{WHITE}{gray}{1}
 \definecolor{RED}{rgb}{1,0,0}
 \definecolor{GREEN}{rgb}{0,1,0}
 \definecolor{BLUE}{rgb}{0,0,1}
 \definecolor{CYAN}{cmyk}{1,0,0,0}
 \definecolor{MAGENTA}{cmyk}{0,1,0,0}
 \definecolor{YELLOW}{cmyk}{0,0,1,0}
 }

\makeatother

\usepackage{babel}
\begin{document}

\title{Quantum deflagration and supersonic fronts of tunneling in molecular magnets}

\author{D. A. Garanin and Saaber Shoyeb}

\affiliation{Department of Physics and Astronomy, Lehman College, City University
of New York, 250 Bedford Park Boulevard West, Bronx, New York 10468-1589,
USA}
\begin{abstract}
Theory of magnetic deflagration taking into account dipolar-controlled
spin tunneling has been applied to the realistic model of molecular
magnet Mn$_{12}$ Ac. At small transverse field, the front speed $v$
has tunneling maxima on the bias field $B_{z}$ reflecting those of
the molecular spin's relaxation rate calculated from the density-matrix
equation. At high transverse field, spin tunneling directly out of
the metastable ground state leads to front speeds that can exceed
the speed of sound. Both for the weak and strong transverse field,
the spatial profile of the deflagration
front near tunneling resonances shows a front of tunneling that triggers a burning front behind it.
\end{abstract}


\maketitle

\section{introduction\label{sec:introduction}}

Burning or deflagration,\cite{gla96book,lanlif9fluid} a self-supporting
phenomenon that can exist in the form of propagating fronts, is decay
of metastable states, controlled by the temperature increasing
as a result of the energy release and heat conduction toward the cold
region before the front. The main ingredient of deflagration is the
decay rate $\Gamma$ of the metastable state that has the Arrhenius
form $\Gamma=\Gamma_{0}\exp\left[-U/(k_{B}T)\right]$ at low temperatures
$T\ll U$, where $U$ is the energy barrier. One could ask if deflagration
can exist in magnetic systems, many of which are bistable due to a
strong uniaxial anisotropy that creates an energy barrier between
the two energy minima. However, the energy release in magnetic systems
is much weaker than in the case of a regular (chemical) deflagration,
thus at room temperatures the ensuing temperature increase is too
small to change the relaxation rate and support burning. The situation
changes at low temperatures, however, since temperature generated
by decay of metastable states can exceed the initial temperature by
far and result in a strong increase of $\Gamma$. Recently magnetic
deflagration has been observed in low-temperature experiments on the
molecular magnet Mn$_{12}$ Ac.\cite{suzetal05prl,heretal05prl} This
discovery initiated theoretical \cite{garchu07prb} and further experimental\cite{mchughetal07prb,hughetal09prb-tuning,hughetal09prb-species}
work. Magnetic deflagration has also been observed on manganites.
\cite{masiaetal07prb} Very fast moving fronts of burning in Mn$_{12}$
Ac initiated by a fast sweep of the magnetic field have been observed
in Ref. \onlinecite{decvanmostejhermac09prl}. This leads to the idea of
magnetic detonation driven by thermal expansion creating a shock wave.\cite{modbycmar11prb,modbycmar11prl}

Main exponents of magnetic deflagration, molecular magnets, are built
of molecules with a large effective spin, such as $S=10$ in Mn$_{12}$
and Fe$_{8}$. Their uniaxial anisotropy $D$ creates the energy barrier
$DS^{2}\simeq67$K for spin rotation\cite{sesgatcannov93nat,baretal96epl}
(see Ref. \onlinecite{gatsesvil06book} for a review). Molecular magnets
made quite a big splash by the discovery of resonance spin tunneling
\cite{frisartejzio96prl,heretal96epl,thoetal96nat} that occurs when
spin energy levels on different sides of the barrier match. This is
controlled by the bias created by the longitudinal magnetic field.
Magnetic molecules in molecular magnets form a crystal lattice (body-centered
tetragonal for Mn$_{12}$ Ac). As magnetic cores of the molecules
are shielded by organic ligands, there is no exchange interaction
between the molecules in the crystal, and the dipole-dipole interaction
(DDI) is dominating. Different members of the Mn$_{12}$ family remain
in the center of magnetic deflagration research because of the elongated
shape of the crystals. To the contrast, Fe$_{8}$ crystals have pyramidal
shape, inappropriate for studying moving fronts.

The impact of spin tunneling on deflagration in molecular magnets
has been addressed in Refs. \onlinecite{heretal05prl,garchu07prb,masiaetal09prb}.
Since no transverse magnetic field was applied in experiments so far,
tunneling via low-lying states was negligibly small. Thus quantum
effects in deflagration could only exist due to thermally assisted
tunneling\cite{chugar97prl,garchu97prb} via the energy levels just
below the top of the barrier. This effect can be taken into account
as effective lowering of the barrier $U$ at resonant values of the
bias.\cite{luisetal97prb} Peaks of the deflagration front speed vs
longitudinal magnetic field (Fig. 4 of Ref. \onlinecite{heretal05prl})
have been interpreted as spin tunneling. The simplest way to explain
these peaks was to use the escape rate $\Gamma$ with the effective
barrier $U$ in the standard formula for the speed of the deflagration
front, Eqs. (\ref{vtil}) and (\ref{Gammaf}) with $\tilde{v}=1$ (dashed line in Fig.
4 of Ref. \onlinecite{heretal05prl}). For higher bias and thinner crystals,
observed speed maxima were much weaker (Fig. 5 of Ref. \onlinecite{mchughetal07prb}
and Fig. 3 of Ref. \onlinecite{hughetal09prb-tuning}) that created a controversy.

At the same time, there was a quest for an essentially quantum mechanism
of deflagration in molecular magnets that does not reduce to mere
barrier lowering in the thermally activated escape rate. As a further
development, fronts of spin tunneling (dubbed {}``cold deflagration'')
controlled by the dipolar field at zero temperature have been proposed.\cite{garchu09prl,gar09prb}
This mechanism requires a strong transverse magnetic field that creates
a sufficiently large tunnel splitting $\Delta$ between the metastable
ground state and an excited state on the other side of the barrier.
The idea is that dipolar field created by the sample produces a bias
on magnetic molecules (spins) that is typically large in comparison
to $\Delta$, thus the dipolar field can control tunneling. As tunneling
of one spin changes dipolar fields on other spins, facilitating or
preventing their tunneling, the problem is self-consistent. It was
shown that there are solutions in which the spatial distribution of
magnetization and dipolar field is adjusting in such a way that there
is a moving front of spin tunneling with many spins in the front core
being on resonance that allows them to tunnel efficiently. This so-called
$laminar$ front has been found for not too large values of the external
bias. For a larger bias it breaks down, resulting in a slow non-laminar
front where most spins are off-resonance.\cite{gar09prb} Fronts of
cold deflagration exist within the dipolar window of the external bias having
the width equal to the dipolar field $B_{z}^{(D)}=52.6$ mT produced
by a uniformly magnetized molecular magnet. \cite{garchu08prb,mchughetal09prb}
In addition to the transverse field, observation of fronts of tunneling
in pure form requires a good thermal contact between the crystal and
its environment, so that released heat gets conducted away and the
temperature remains low.

If the crystal of a molecular magnet is thermally insulated, spin
tunneling in a biased case leads to release of Zeeman energy and the
temperature increase. In this case both spin tunneling and thermal
activation can play a role, so that deflagration is controlled by
two parameters, dipolar field and the temperature. The combined quantum-thermal
theory of magnetic deflagration has been proposed in Ref. \onlinecite{garjaa10prbrc}.
In contrast to the pure cold deflagration, where in the case of overdamped
tunneling it is sufficient to use the Lorentzian form of the tunneling
rate near the resonance {[}Eq. (12) of Ref. \onlinecite{gar09prb}{]}, here
one needs the numerically calculated escape rate $\Gamma(B_{z},T)$
for both resonant and non-resonant values of $B_{z}$. This escape
rate has been calculated from the density matrix equation \cite{gar11acp}
based on the universal spin-phonon interaction. \cite{chu04prl,chugarsch05prb}
To the contrast with the pure cold deflagration that leaves some metastable
magnetization unburned behind the front, the combined deflagration
leads to complete burning, as the standard magnetic deflagration.
This flattens out irregularities of non-laminar fronts and makes them
move faster, reaching high speeds at the right end of the dipolar
window (see Fig. 4 of Ref. \onlinecite{garjaa10prbrc}).

Ref. \onlinecite{garjaa10prbrc} used the generic model of a molecular magnet
with the anisotropy of the form $-DS_{z}^{2}$. In this model, tunneling
resonances of all levels take place at the same value of $B_{z}$
\begin{equation}
B_{z}=B_{k}=kD/(g\mu_{B}),\qquad k=0,\pm1,\pm2,\ldots\label{Bk}
\end{equation}
and, non-trivially, the resonances remain unchanged if transverse
magnetic field is applied. In the real Mn$_{12}$ Ac there is an additional
term $-AS_{z}^{4}$ that makes resonances of different levels be achieved
at different values of $B_{z}$. The latter was used to experimentally
monitor the transition between thermally assisted and ground-state
tunneling in Mn$_{12}$ Ac.\cite{bokkenwal00prl,wermugchr06prl} Splitting
of tunneling resonances should manifest itself in experiments on magnetic
deflagration, and studying related phenomena is one of the aims of
this work.

Another aim of this work is to explore the high-speed regime of magnetic
burning near the ground-state resonance at high transverse fields.
As the speed of fronts of tunneling should be much higher than that
of the standard burning fronts driven by heat conduction, burning
in these fronts should be independent of the thermal diffusivity,
that resembles detonation. To study this regime, more accurate numerical
calculations on longer crystals have to be performed.

The rest of the paper is organized as follows. In Sec. \ref{sec:Equations-of-deflagration}
equations describing deflagration with dipolar-controlled spin tunneling
are set up and the method of their solution is outlined. Sec. \ref{sec:The-relaxation-rate}
introduces the relaxation rate of magnetic molecules that is calculated
with the help of the density-matrix formalism and contains the effects
of both thermal activation and spin tunneling. Sec. \ref{sec:Front-speed-weak-field}
presents numerical results for the front speed in weak transverse
fields. Sec. \ref{sec:Front-speed-strong-field} is devoted to the
case of a strong transverse field, where ground-state tunneling leads
to supersonic front speeds. Concluding section summarizes the results
obtained and outlines unsolved problems.

\section{Equations of deflagration with spin tunneling and dipolar field\label{sec:Equations-of-deflagration}}

The system of equation describing deflagration with quantum effects
in molecular magnets\cite{garjaa10prbrc} consists of the rate equation
for the metastable population $n$
\begin{equation}
\frac{\partial n(t,z)}{\partial t}=-\Gamma\left(B_{\mathrm{tot},z}(z),T(z)\right)\left[n(t,z)-n^{(\mathrm{eq})}(T)\right]\label{ndot}
\end{equation}
and the heat conduction equation that can be conveniently written
for the thermal energy $\mathcal{E}$ per magnetic molecule
\begin{equation}
\frac{\partial\mathcal{E}(t,z)}{\partial t}=\frac{\partial}{\partial z}\kappa\frac{\partial\mathcal{E}(t,z)}{\partial z}-\Delta E\frac{\partial n(t,z)}{\partial t}.\label{Edot}
\end{equation}
It is assumed that the crystal has an elongated shape and everything
depends only on the coordinate $z$ along the geometrical axis of
the crystal. The easy axes of magnetic molecules are also directed
along this axis, that was the case for all experimentally studied
crystals.\cite{suzetal05prl,heretal05prl,mchughetal07prb,mchughetal09prb,hughetal09prb-species,hughetal09prb-tuning}
In Eq. (\ref{ndot}) $\Gamma(B_{z},T)$ is the numerically computed
relaxation (escape) rate of magnetic molecules' spins out of the metastable
state with the spin pointed to the left, when a longitudinal external
field is applied in the direction to the right. $n^{(\mathrm{eq})}(T)$
is the thermal-equilibrium population of the metastable state that
is small in the case of a large bias and will be discarded. In Eq.
(\ref{Edot}) $\kappa$ is thermal diffusivity that proves to be difficult
to measure. Estimations\cite{suzetal05prl} yield $\kappa\sim10^{-5}$m$^{2}$/s
(comparable to that of metals) that will be adopted here. The second
term in this equation is the source term, in which $\Delta E$ is
the energy released by transition of one molecular spin from the metastable
state to the ground state $\left|-S\right\rangle \rightarrow\left|S\right\rangle $,
that is, $\Delta E=2Sg\mu_{B}B_{z}$. The relation between the energy
$\mathcal{E}$ and temperature is given by
\begin{equation}
\mathcal{E}(T)=\int_{0}^{T}C(T^{\prime})dT^{\prime},\label{EviaT}
\end{equation}
 where $C(T)$ is the experimentally measured heat capacity of Mn$_{12}$Ac
per magnetic molecule.\cite{gometal98prb}

Since the relaxation rate $\Gamma(B_{\mathrm{tot},z},T)$ has very
sharp maxima at the resonance values of the total longitudinal field
$B_{\mathrm{tot},z}$, it is important to include the dipolar field
created by the crystal,
\begin{equation}
B_{\mathrm{tot},z}(z)=B_{z}+B_{z}^{(D)}(z).\label{Btotz}
\end{equation}
Although the dipolar field $B_{z}^{(D)}$ is much weaker than the
external field $B_{z}$ (and thus can be dropped in $\Delta E$),
it is much greater than the width of tunneling peaks in $\Gamma(B_{\mathrm{tot},z},T)$,
so that it can control tunneling. It is convenient to represent $B_{z}^{(D)}$
in the form
\begin{equation}
B_{z}^{(D)}=\frac{Sg\mu_{B}}{v_{0}}D_{zz},\label{BviaD}
\end{equation}
where $D_{zz}$ is the dimensionless dipolar field, $v_{0}=a^{2}c$
is the unit-cell volume, $a$ and $c$ are lattice spacings. For Mn$_{12}$
Ac one has $Sg\mu_{B}/v_{0}=5.0$ mT. For crystals of cylindrical
shape with radius $R$ and length $L$ one obtains\cite{garchu08prb}
\begin{equation}
D_{zz}(z)=\int_{0}^{L}dz^{\prime}\frac{2\pi\nu R^{2}\sigma_{z}(z^{\prime})}{\left[\left(z^{\prime}-z\right)^{2}+R^{2}\right]^{3/2}}-k_{D}\sigma_{z}(z),\label{DzzCylinh}
\end{equation}
where $\nu$ is the number of molecules per unit cell, $\nu=2$ for Mn$_{12}$
Ac, $\sigma_{z}=1-2n$ is polarization of pseudospins representing
spins of magnetic molecules ($\sigma_{z}=\pm1$ in the ground and
metastable states, respectively) and
\begin{equation}
k_{D}\equiv8\pi\nu/3-\bar{D}_{zz}^{(\mathrm{sph})}=4\pi\nu-\bar{D}_{zz}^{(\mathrm{cyl})}>0,\label{kDef}
\end{equation}
Here the barred quantities correspond to the reduced dipolar field
inside a uniformly magnetized sphere and a long cylinder, and $D_{zz}=\bar{D}_{zz}\sigma_{z}$
for $\sigma_{z}=\mathrm{const}$. For Mn$_{12}$ Ac calculations yield\cite{garchu08prb}
$\bar{D}_{zz}^{(\mathrm{sph})}=2.155$, $\bar{D}_{zz}^{(\mathrm{cyl})}=10.53$
(in real units $B_{z}^{(D)}=52.6$ mT \cite{garchu08prb,mchughetal09prb}),
and thus in the local term of Eq. (\ref{DzzCylinh}) one has $k_{D}=14.6$.
One can check that Eq. (\ref{DzzCylinh}) yields the correct result
for the field inside a long uniformly magnetized cylinder. At the
ends of a cylinder dipolar field has the form $D_{zz}=\left(\bar{D}_{zz}^{(\mathrm{sph})}-2\pi\nu/3\right)\sigma_{z}$
that for Mn$_{12}$ Ac becomes $D_{zz}=-2.03\sigma_{z}$. Dipolar
field opposite to the spin orientation is the reason for the instability
of the uniformly magnetized state of Mn$_{12}$ Ac that leads to domain
formation.\cite{gar10prbrc} For other shapes such as elongated rectangular,
one obtains qualitatively similar expressions.\cite{gar09prb}

It
has to be stressed that the results above represent the dipolar field
exactly at the magnetic molecules in the lattice and they depend on
the lattice structure. Using the spatially averaged field following
from macroscopic magnetostatics would be a mistake.
Indeed, the magnetostatic field inside a long uniformly magnetized cylinder is ${\cal B}_{z}^{(D)}=4\pi M$, where the magnetization
is given by $M=\nu Sg\mu_{B}/v_{0}$. The dipolar field in Mn$_{12}$ Ac is essentially smaller,
$B_{z}^{(D)}=(\bar{D}_{zz}^{(\mathrm{cyl})}/2) M = 5.26 M$.
The difference between the two is due to the local term with $k_D$ in Eq. (\ref{DzzCylinh}).

\begin{figure}
\includegraphics[angle=-90,width=8cm]{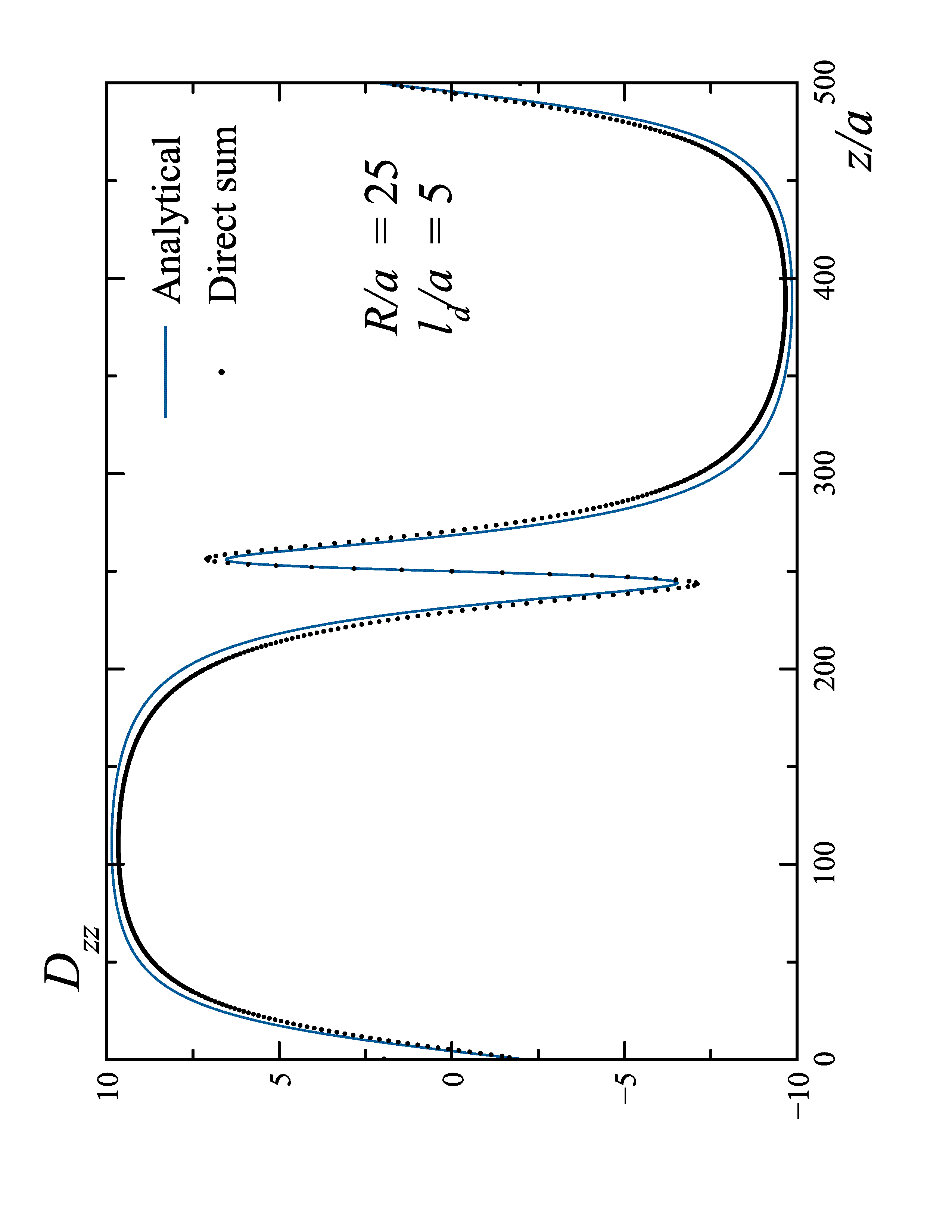}

\caption{Reduced dipolar field in a deflagration front in the slow-burning
limit, created by the magnetization profile $\sigma_{z}(z)=-\tanh\left[(z-z_{0})/l_{d}\right]$.
Solid line: Eq. (\ref{DzzCylinh}); Points: Direct summation of dipolar
fields over the Mn$_{12}$ Ac lattice. }
\label{Fig-Dzz-profile}
\end{figure}

A striking feature of Eq. (\ref{DzzCylinh}) is that the integral
and local terms have different signs. The integral term changes at
the scale of $R$ while the local term can change faster, that creates
a non-monotonic dependence of $D_{zz}(z)$. In the case of a regular
magnetic deflagration, the spatial magnetization profile in the slow-burning
limit is $\sigma_{z}(z)=-\tanh\left[(z-z_{0})/l_{d}\right]$, where
$l_{d}$ is the width of the deflagration front that satisfies $l_{d}\ll R$
(see below). The resulting dipolar field is shown in Fig. \ref{Fig-Dzz-profile},
where the line is the result of Eq. (\ref{DzzCylinh}) and points
represent the dipolar field along the symmetry axis of a long cylindrical
crystal calculated by direct summation of microscopic dipolar fields
over the Mn$_{12}$ Ac lattice. One can see that Eq. (\ref{DzzCylinh})
is pretty accurate, small discrepancies resulting from $l_{d}$ being
not large enough in comparizon to the lattice spacing $a.$ The central
region with the large positive slope is dominated by the local term
of Eq. (\ref{DzzCylinh}) that changes in the direction opposite to
that of the magnetization. For $R\ggg l_{d}$, $D_{zz}$ reaches the
values $\pm14.6$ due to the local term before it begins to slowly
change in the opposite direction. In real units the dipolar field
at the local maximum and minimum is $\pm B_{z}^{(k_{D})}$, where
\begin{equation}
B_{z}^{(k_{D})}=72.9\,\mathrm{mT}\label{BzkD}
\end{equation}
exceeding the dipolar field of the uniformly magnetized long cylinder
$B_{z}^{(D)}=52.6$ mT. Also one can see from Fig. \ref{DzzCylinh}
that the dipolar field becomes opposite to the magnetization at the
ends of the cylinder, as mentioned above.

Equations (\ref{ndot})--(\ref{DzzCylinh}) form a system of integro-differential
equations describing deflagration with spin tunneling in molecular
magnets taking into account the dipole-dipole interaction. Before
discussing numerical solution of these equations, it is worth recuperating
the results of the standard ({}``hot'') deflagration and of the
cold deflagration. If the whole released energy remains in the body
and the initial temperature is very low, the thermal energy per spin
behind the front is $\Delta E$. The corresponding temperature defined
by the inversion of Eq. (\ref{EviaT}) is the so-called flame temperature
$T_{f}=T(\Delta E)$ that is in the range 10--15 K in deflagration
experiments. Theory of deflagration yields the expressions for the
speed of the front $v$ and front width $l_{d}$
\begin{equation}
v=\tilde{v}\Gamma_{f}l_{d}=\tilde{v}\sqrt{\kappa_{f}\Gamma_{f}},\qquad l_{d}=\sqrt{\kappa_{f}/\Gamma_{f}},\label{vtil}
\end{equation}
where $\kappa_{f}$ and
\begin{equation}
\Gamma_{f}=\Gamma_{0}\exp\left(-W_{f}\right),\qquad W_{f}\equiv U/(k_{B}T_{f})\label{Gammaf}
\end{equation}
are thermal diffusivity and relaxation rate at the flame temperature,
while $\tilde{v}$ is a dimensionless coeficient. It was shown\cite{garchu07prb}
that in the slow-burning limit $W_{f}\gg1$ one has $\tilde{v}\cong2/\sqrt{W_{f}}$.
On the other hand, the speed of the laminar front of tunneling at
zero temperature is given by\cite{garchu09prl,gar09prb}
\begin{equation}
v=v^{*}\Gamma_{\mathrm{res}}R,\label{vstar}
\end{equation}
where $\Gamma_{\mathrm{res}}=\Delta^{2}/(\hbar^{2}\Gamma_{m'})$ is
the relaxation rate at overdamped tunneling resonance, $\Delta/\hbar\lesssim\Gamma_{m'}$,
$\Gamma_{m'}$ being the decay rate of the matching level $m'$ at
the other side of the barrier, $R$ is the width of the crystal (radius
of the cylinder in our model), and $v^{*}$ is a dimensionless coefficient.
With a sufficiently strong transverse field applied, one can have
$\Delta/\hbar\sim\Gamma_{m'}$ at the applicability limit of the overdamped
approximation, and then $\Gamma_{\mathrm{res}}\gg\Gamma_{f}$ because
thermal activation goes over high levels of the magnetic molecule
where the distances between the levels and thus the energies of phonons
involved are much smaller than for the low-lying levels, and also
because $\Gamma_{f}$ is exponentially small since $T_{f}\lesssim U$.
Additionally, estimation of $l_{d}$ with $\kappa_{f}=10^{-5}$m$^{2}$/s
and the experimental value $\Gamma_{0}=10^{7}$s$^{-1}$ yield $l_{d}\sim3\times10^{-4}$
mm for $B_{z}$ near the first tunneling resonance and even smaller
for larger bias. As in the experiment the width of the crystal was
much larger than $l_{d}$ (0.3 mm in Ref. \onlinecite{suzetal05prl}, 0.2
mm in Ref. \onlinecite{mchughetal07prb} and 1 mm in Ref. \onlinecite{heretal05prl}),
one can see that $\Gamma_{\mathrm{res}}R\gg\Gamma_{f}l_{d}$ is quite
possible in a strong transverse field, and then the front of spin
tunneling is much faster than the front of spin burning. A very conservative
estimation with $\Gamma_{\mathrm{res}}\Rightarrow\Gamma_{0}=10^{7}$
s$^{-1}$ and $v^{*}\Rightarrow1$ for the crystal 0.2 mm thick yields
$v\sim1000$ m/s. As said above, in a strong transverse field one
can have $\Gamma_{\mathrm{res}}\gg\Gamma_{0}$, so that the speed
of a spin-tunneling front can easily surpass the speed of sound that
is about 2000 m/s in molecular magnets (see analysis in Ref. \onlinecite{gar08prbrc}).
The results of our calculations confirm this.

Discretization of the variable $z$ reduces Eqs. \ref{ndot}, \ref{Edot},
and \ref{DzzCylinh} to a system of ordinary differential equations
that can be solved numerically. Very narrow tunneling peaks in $\Gamma(B_{z},T)$
make it necessary to carefully control the step in the numerical integration.
Wolfram Mathematica's NDSolve proves to be an efficient tool for this
problem. To ignite a deflagration front, the temperature at the left
end of the crystal had been increased during a short time. Then the
equations were solved and, to find the front speed, the time of arrival
of the front at the right end of the crystal was
measured.

\section{The relaxation rate\label{sec:The-relaxation-rate}}

\begin{figure}
\includegraphics[angle=-90,width=8cm]{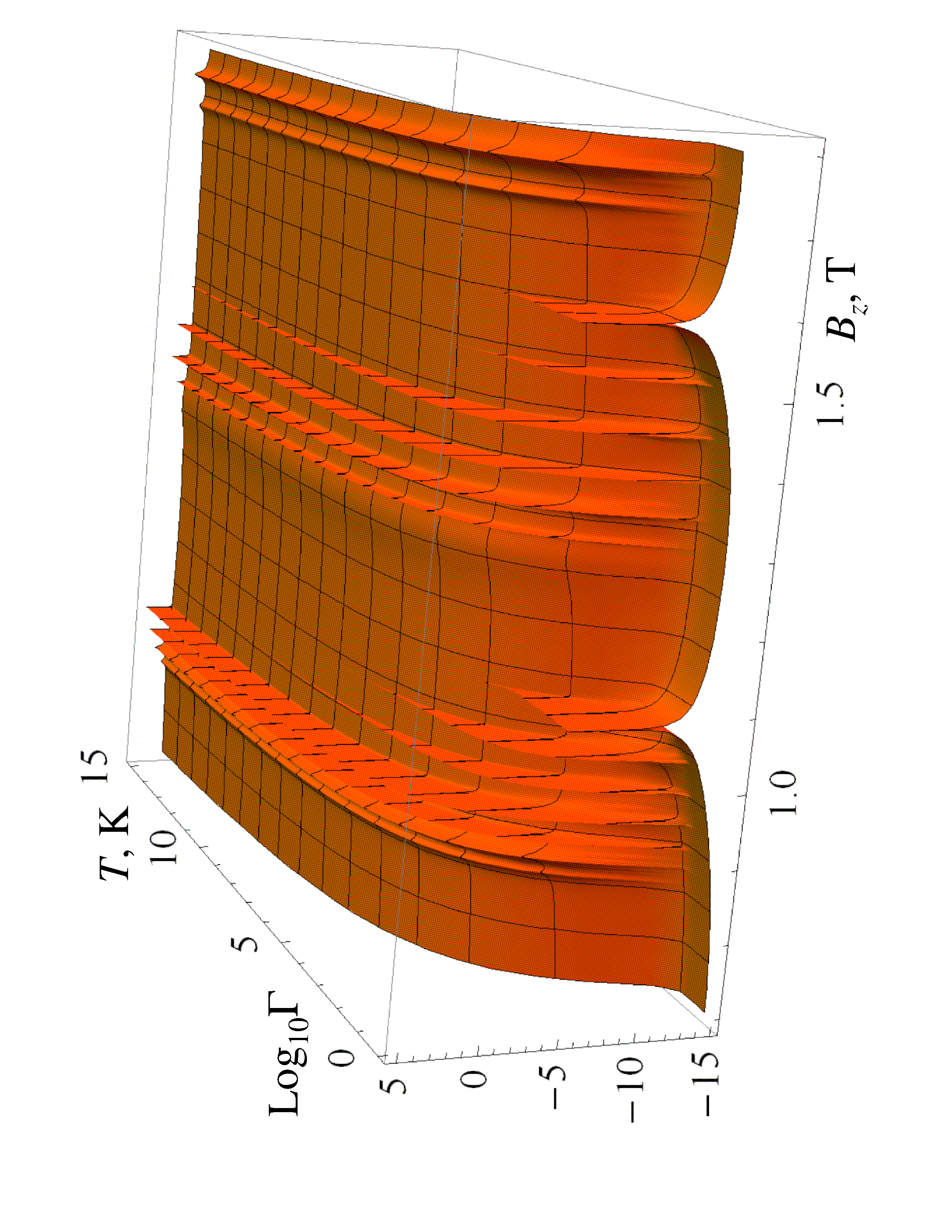}

\caption{Relaxation rate of Mn$_{12}$Ac vs temperature and longitudinal magnetic
field in the transverse field $B_{\bot}=0.04$ T.}
\label{Fig-3D-Gamma_Javier}
\end{figure}

\begin{figure}
\includegraphics[angle=-90,width=8cm]{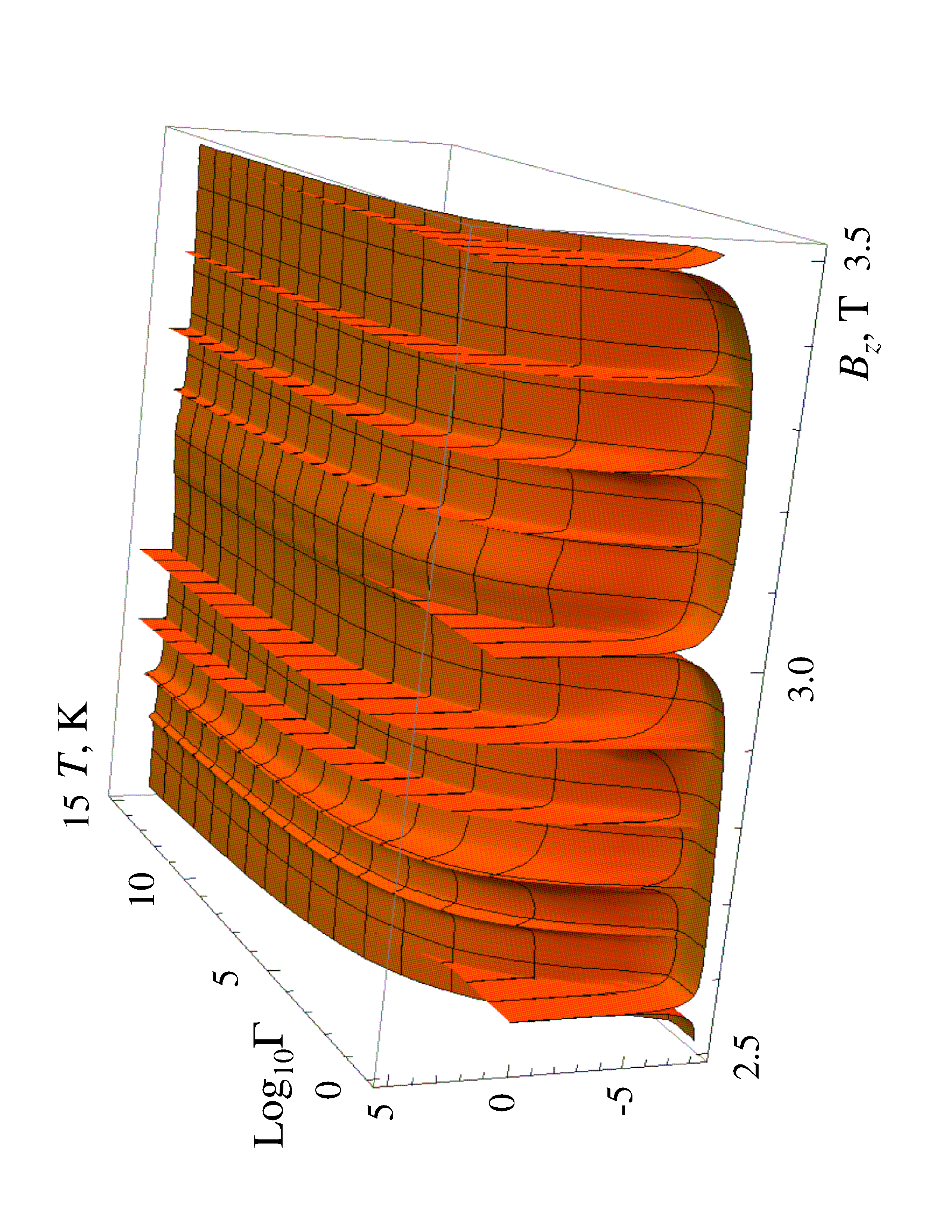}

\caption{Relaxation rate of Mn$_{12}$Ac vs temperature and longitudinal magnetic
field in the transverse field $B_{\bot}=0.04$ T for a stronger bias.}
\label{Fig-3D-Gamma_Myriam}
\end{figure}

It is crucial to calculate and tabulate the relaxation rate $\Gamma(B_{z},T)$
before solving the deflagration problem because a runtime calculation
of $\Gamma(B_{z},T)$ is practically impossible. We use the effective-spin model
with the Hamiltonian containing the uniaxial anisotropy $-DS_{z}^{2}-AS_{z}^{4}$
and other anisotropy terms, according to Ref. \onlinecite{barkenrumhencri03prl}.
Spin-phonon interaction is taken into account within the universal
model of pure rotations of the crystal field by transverse phonons
described in Refs. \onlinecite{chu04prl,chugarsch05prb,calchugar06prb}.
Since in this model the crystal field is not distorted, spin-phonon
coupling coefficients can be expressed through the measurable crystal-field
parameters. The density-matrix equation has been solved within the
semi-secular approximation that is valid everywhere, including tunneling
resonances.\cite{gar11acp}

In the generic model of a molecular magnet with the anisotropy $-DS_{z}^{2}$
the fields corresponding to tunneling resonances are given by Eq. (\ref{Bk})
for all level pairs. The resulting $\Gamma(B_{z},T)$ in a strong
transverse field is shown in Fig. 2 of Ref. \onlinecite{garjaa10prbrc}.
Tabulation of such a function requires a lot of points along the $B_{z}$
axis in the vicinity of tunneling maxima. The realistic model with
the uniaxial anisotropy $-DS_{z}^{2}-AS_{z}^{4}$ is more complicated
because tunneling resonances for different level pairs are achieved
for different $B_{z}$ that depend on the transverse field. Thus the
first step is to find tunneling peaks numerically for a given transverse
field, then to build a non-equidistant grid with a small step near
the peaks, then calculate $\Gamma(B_{z},T)$ and, finally, make the
interpolation. These tasks have been fulfilled with the help of Wolfram
Mathematica using a high custom precision and parallelization.

For a weak transverse field (set to $B_{\bot}=0.04-0.05$ T that may
result from a 1\textdegree{} misalignment between the crystal axis
and the longitudinal field) $\Gamma(B_{z},T)$ contains a zoo of tunneling
peaks shown in Fig. \ref{Fig-3D-Gamma_Javier}. The range of $B_{z}$
here corresponds to that in Ref. \onlinecite{heretal05prl} and contains
groups of resonances with $k=2,3$ and partially 4. One can see that
ground-state resonances, that are the only survivors at $T=0$, are
achieved at higher fields than resonances of excited states. At temperatures
as high as flame temperature, low-lying tunneling resonances are drowned
in the non-resonant background. There is also a much weaker non-resonant
tunneling at $T=0$. Relaxation rate at a stronger bias, also in a small
transverse field, corresponding to that in Refs. \onlinecite{suzetal05prl,mchughetal07prb,hughetal09prb-tuning,hughetal09prb-species},
is shown in Fig. \ref{Fig-3D-Gamma_Myriam}. At such bias, the effect
of ground-state tunneling begins to appear at high temperatures.

\begin{figure}
\includegraphics[angle=-90,width=8cm]{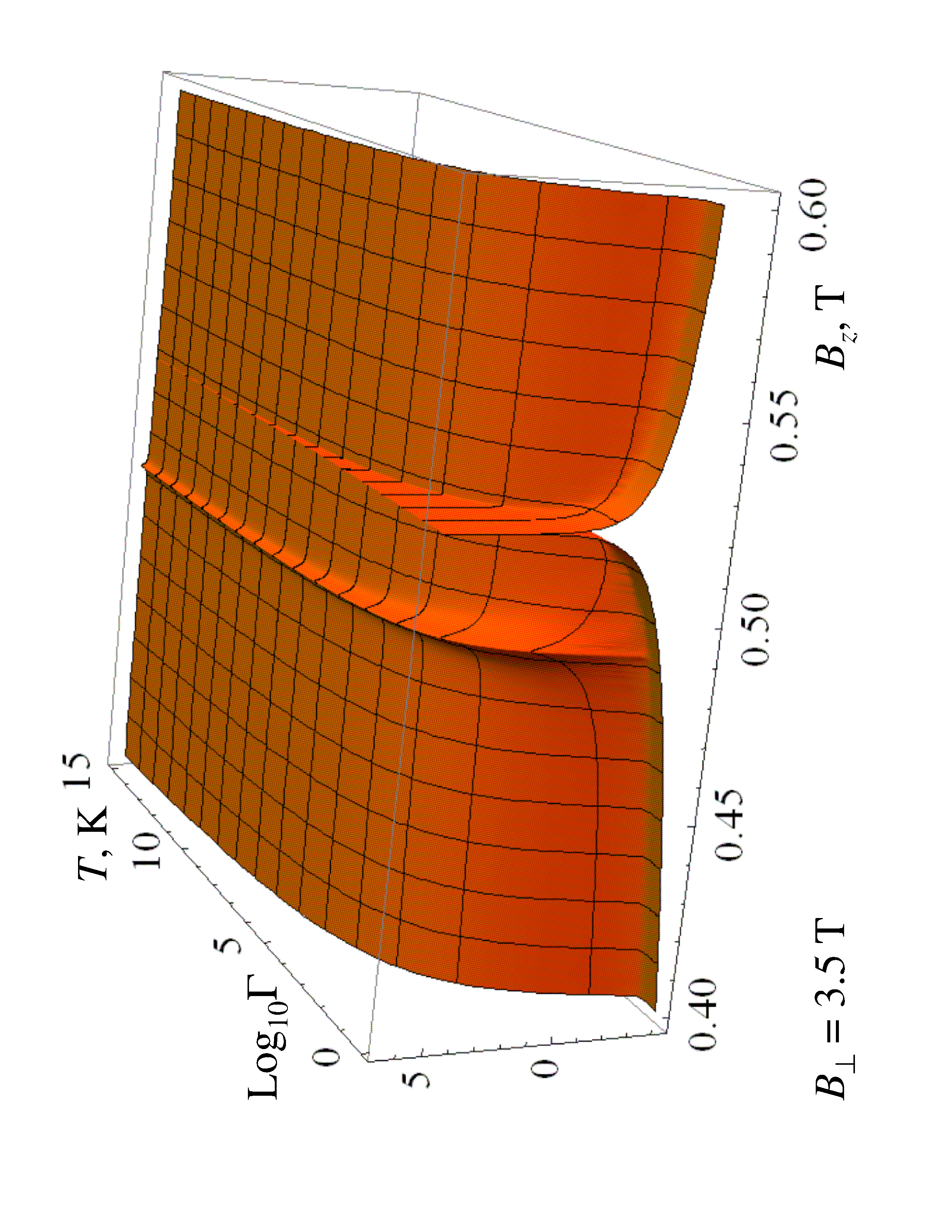}

\caption{Relaxation rate of Mn$_{12}$Ac vs temperature and longitudinal magnetic
field in the transverse field $B_{\bot}=3.5$ T.}
\label{Fig-3D-Gamma_Btr=00003D3.5T}
\end{figure}

\begin{figure}
\includegraphics[angle=-90,width=8cm]{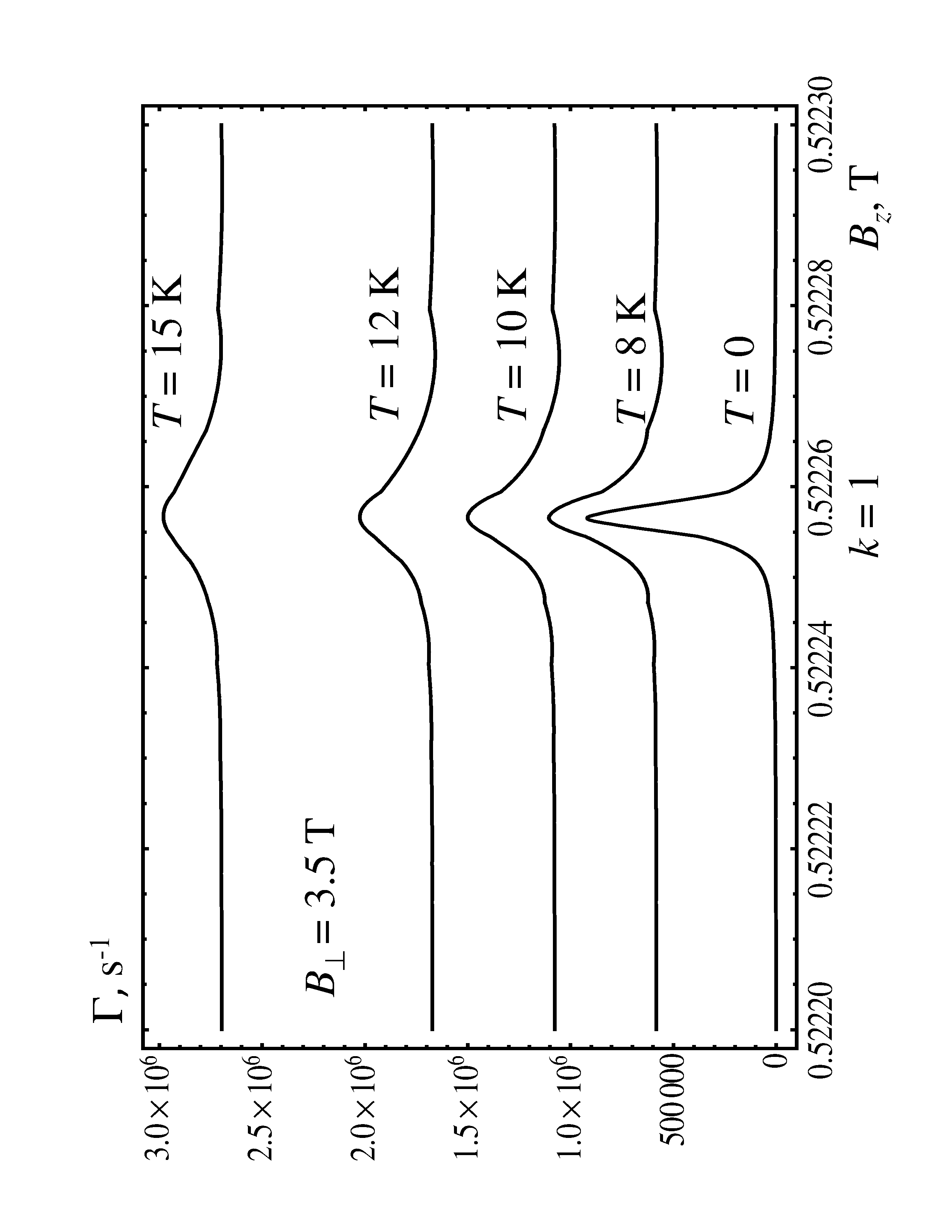}

\caption{Magnification of the ground-state tunneling peak of $\Gamma(B_{z},T)$ (multiplied by 100) at $B_{\bot}=3.5$
T. }
\label{Fig-Gamma-Btr=00003D3.5T-ground-state_peak}

\end{figure}

In a strong transverse field such as $B_{\bot}=3.5$ T in Fig. \ref{Fig-3D-Gamma_Btr=00003D3.5T}
the barrier is strongly lowered and most of tunneling resonances are
broadened away. Here one can see the ground-state resonance ($B_{z}=0.522$),
the first-excited-state resonance ($B_{z}=0.490$), and with an effort
a very broad second-excited-state resonance further to the left. Note
the much higher tunneling rate at $T=0$, in comparison with the previous
figure. The range of $B_{z}$ in Fig. \ref{Fig-3D-Gamma_Btr=00003D3.5T}
corresponds to the tunneling resonance with $k=1$. Fig. \ref{Fig-Gamma-Btr=00003D3.5T-ground-state_peak}
shows the details of the ground-state peak in Fig. \ref{Fig-3D-Gamma_Btr=00003D3.5T}.
The height and width of this peak increase with temperature. This
increase is moderate, however, in comparison to the exponential increase
of the non-resonant relaxation rate. The first-excited-state peak
in Fig. \ref{Fig-3D-Gamma_Btr=00003D3.5T} is higher than the ground-state
peak at the flame temperature but is plays a much smaller role in
the front propagation, as we will see below.

\begin{figure}
\includegraphics[angle=-90,width=8cm]{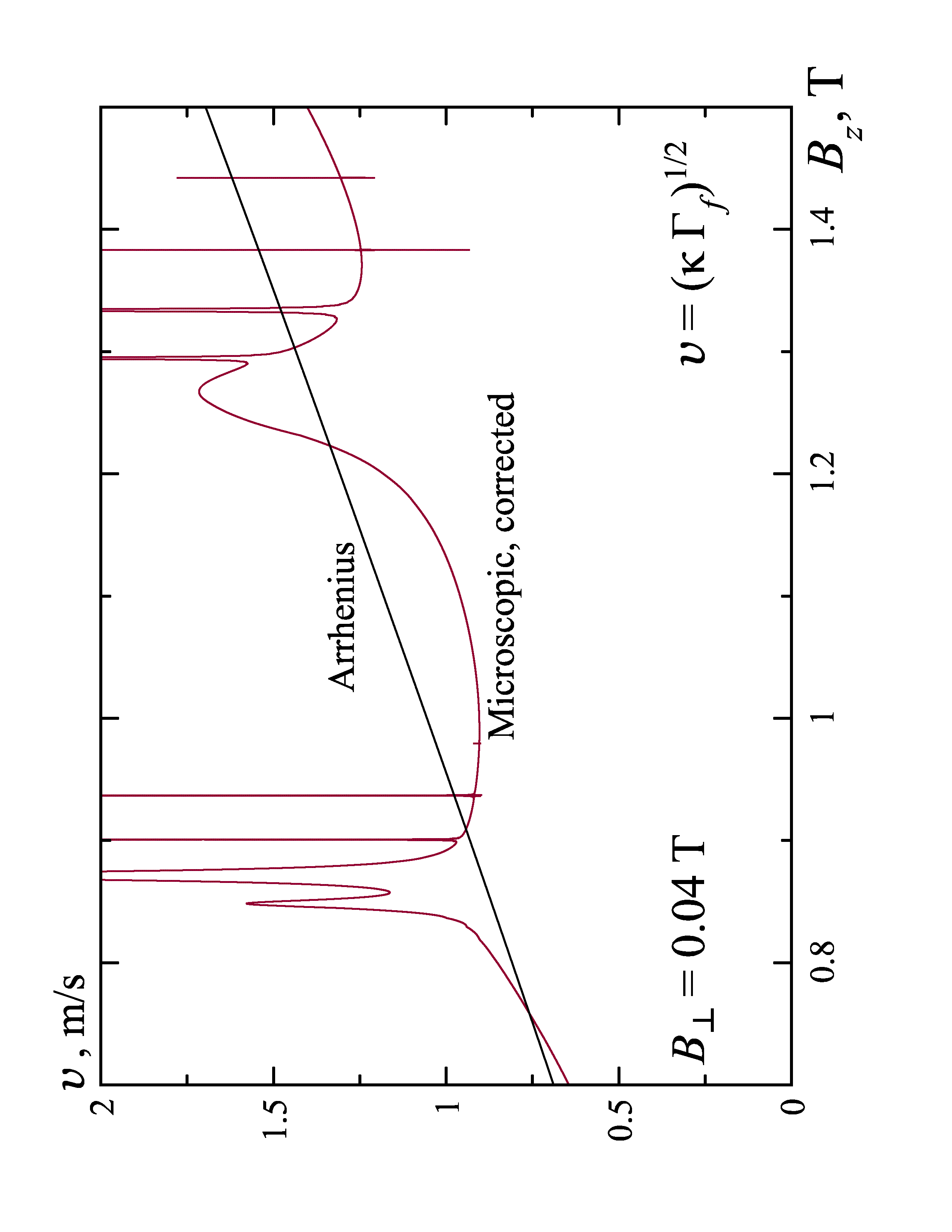}

\caption{Speed of deflagration front estimated from Eq. (\ref{vtil}) for the
microscopically calculated relaxation rate $\Gamma(B_{z},T_{f}(B_{z}))$
with the correction factor 100, together with Eq. (\ref{vtil}) using
the Arrhenius formula for $\Gamma(B_{z},T_{f}(B_{z}))$.}
\label{Fig-v_Estimation_BzExt=00003D0.7-1.5T}

\end{figure}

A long-standing problem in the theory of relaxation of molecular magnets
is the prefactor $\Gamma_{0}$ in
the Arrhenius relaxation rate being by two orders of magnitude too small.
This was already recognized in the
early Ref. \onlinecite{garchu97prb}. Without introducing artificially strong
spin-phonon interactions, \cite{leulos99epl} it is impossible to
arrive at $\Gamma_{0}\sim10^{7}$s$^{-1}$ observed in experiments\cite{luisetal97prb,gometal98prb}
using the standard spin-lattice relaxation model considering one spin
in an infinite elastic matrix. This model could be justified for a
strongly diluted molecular magnet but in the normal case it cannot.
High density of magnetic molecules should lead to such collective
effects as superradiance\cite{dic54,chugar02prl,chugar04prl} and
phonon bottleneck.\cite{abrble70,gar07prb,gar08prb} As it would be
difficult to deal with these complicated issues while addressing the
quantum deflagration problem, the calculated relaxation
rate was simply multiplied by 100 to approximately match the experiment. It is instructive
to plot the theoretical deflagration speed given by Eq. (\ref{vtil})
(with $\tilde{v}=1$) at small transverse field as function of $B_{z}$
using the corrected values of $\Gamma(B_{z},T_{f}(B_{z}))$. Fig.
\ref{Fig-v_Estimation_BzExt=00003D0.7-1.5T} shows a good overall
agreement, except for tunneling maxima in the microscopically calculated
result. As tunneling resonances are broadened by ligand disorder,
dipolar field, and nuclear spins, very narrow peaks due to tunneling
resonances of lower levels here will be washed out in the experimentally
measured front speed. In fact, a similar interpretation of experimental
results have been done in Ref. \onlinecite{heretal05prl}, where the dashed
line in Fig. 4 is $\sqrt{\kappa_{f}\Gamma_{f}}$ with $\Gamma_{f}$
taken from relaxation experiments on the same crystal.

An alternative explanation of much higher relaxation rates observed in the experiment is based on deviations from
the strong-exchange model that lead to mixing of the states with different total spin $S$.
In Ref. \onlinecite{caretal04prl} it was shown that this small mixing taken into account perturbatively leads again
to the giant-spin model with $S=10$, however, with additional higher-order crystal-field terms that would normally
be absent for $d$-electrons.
These additional terms can explain the observed ground-state tunnel splitting $\Delta$ in Fe$_8$ that is three orders of magnitude
larger than the theoretical result using the standard spin Hamiltonian.
Similar mechanism could work for Mn$_{12}$ and lead to the increase of the spin-lattice relaxation rate as well.
However, the importance of this mechanism is limited to small transverse fields.
Most interesting results below for supersonic fronts of tunneling directly out of the metastable ground state without
thermal activation require a strong transverse field that produces a large tunnel splitting.
In this limit, the latter becomes insensitive to crystal-field terms responsible for tunneling in zero or small transverse fields.

\section{Front speed at weak transverse field\label{sec:Front-speed-weak-field}}

\begin{figure}

\includegraphics[angle=-90,width=8cm]{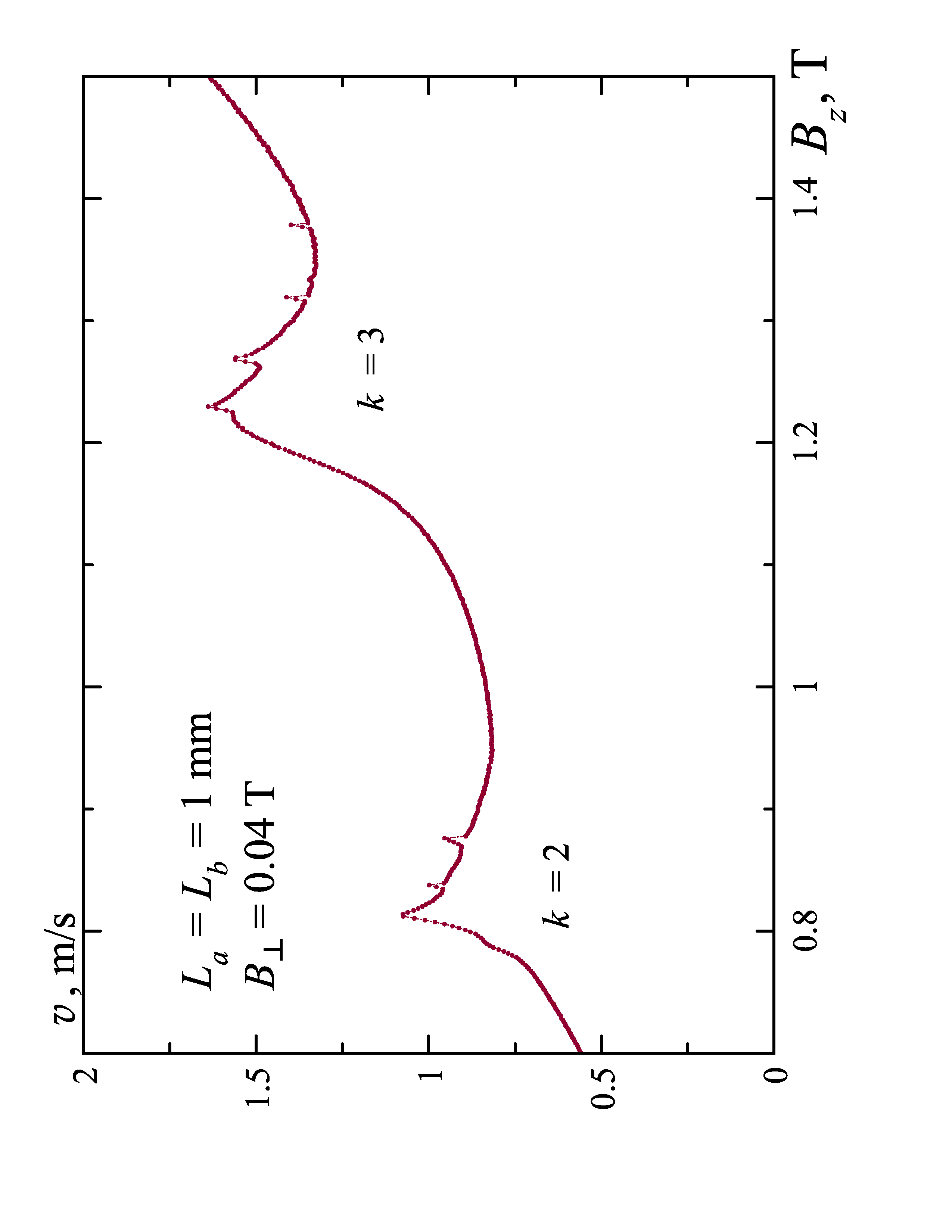}\caption{Numerically calculated speed of the deflagration front in a Mn$_{12}$
Ac crystal in small transverse field. }
\label{Fig-v_Bz=00003D0.7-1.6T-Javie}

\end{figure}

\begin{figure}
\includegraphics[angle=-90,width=8cm]{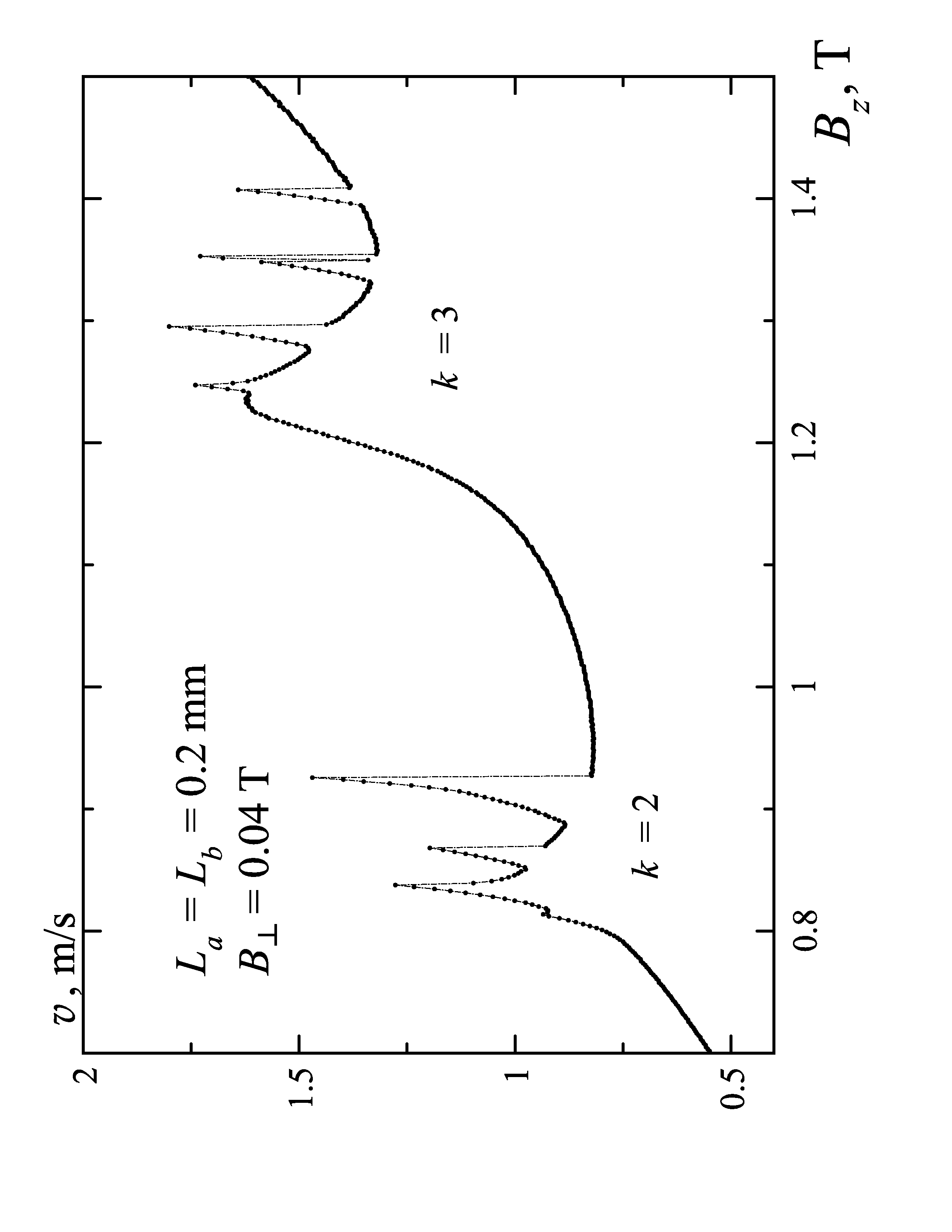}

\caption{Numerically calculated speed of the deflagration front for small transverse
field for a Mn$_{12}$ Ac crystal with a smaller transverse size.}

\label{Fig-v_Bz=00003D0.7-1.6T-Myriam}
\end{figure}

\begin{figure}
\includegraphics[angle=-90,width=8cm]{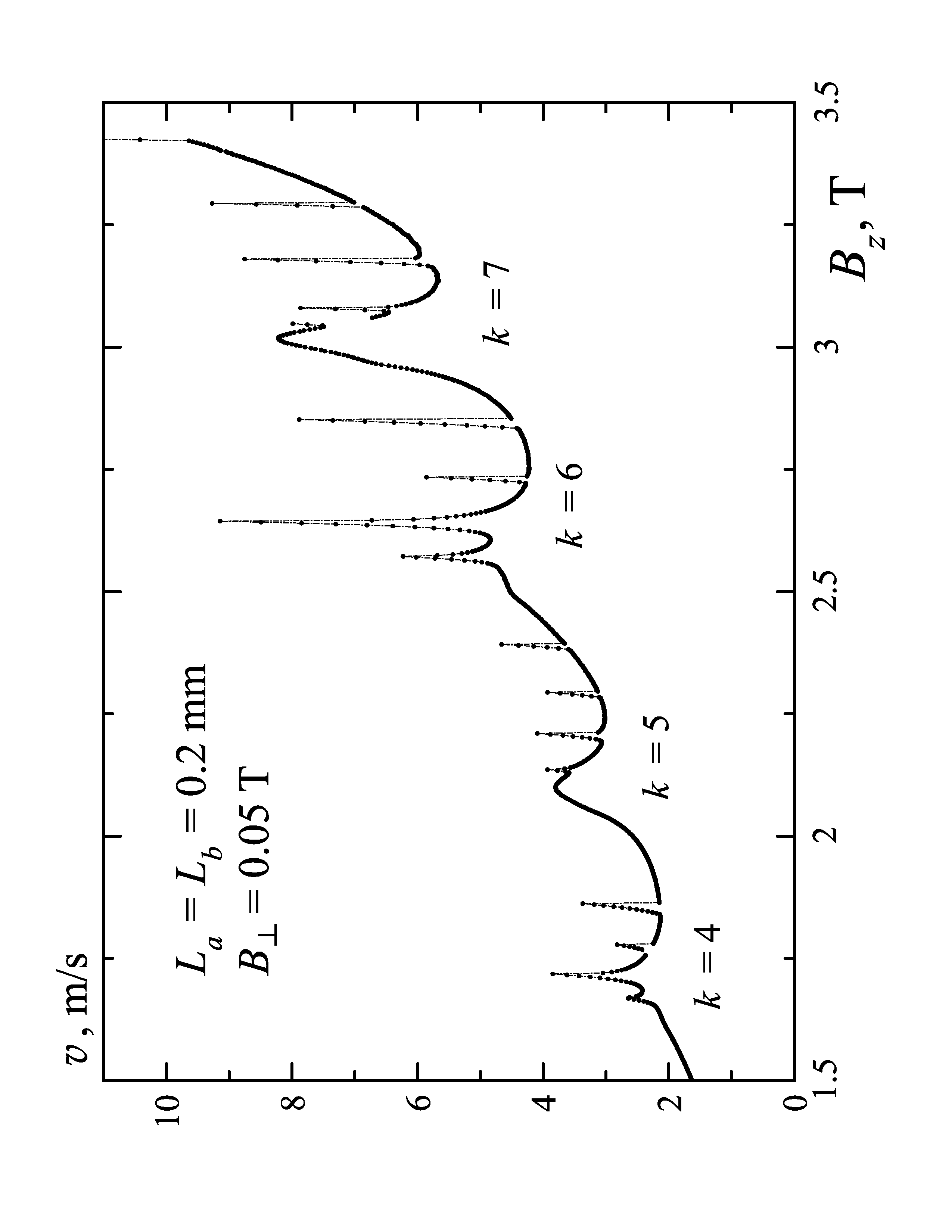}

\caption{Numerically calculated speed of the deflagration front in a thinner
Mn$_{12}$ Ac crystal for small transverse field and a larger bias.}
\label{Fig-v_Bz=00003D1.5-3.5T}

\end{figure}

The procedure of numerical solution of the quantum deflagration equations
is discussed at the end of Sec. \ref{sec:introduction}. The result
for the front speed at small transverse fields in the range $B_{z}=$0.7--1.7
T is shown in Fig. \ref{Fig-v_Bz=00003D0.7-1.6T-Javie}. Here the
cylinder radius $R$ in our model has been chosen so that it yields
the same cross-section as the crystal of transverse sizes $L_{a}=L_{b}=1$
mm in Ref. \onlinecite{heretal05prl}, that is, $R=\sqrt{L_{a}L_{b}/\pi}=0.564$
mm. One can see that, in comparison to Fig. \ref{Fig-v_Estimation_BzExt=00003D0.7-1.5T},
narrow tunneling peaks are washed out and only broad peaks remain.
The reason is that the total magnetic field in the crystal is not
constant and changes in the front as shown in Fig. \ref{Fig-Dzz-profile},
so that tunneling resonances in $v$ are spread.  Overall there is
a good agreement between our Fig. \ref{Fig-3D-Gamma_Javier} and Fig.
4 of Ref. \onlinecite{heretal05prl}.
For a comparizon,
the calculated front speed for a crystal of smaller
transverse dimensions, $L_{a}=L_{b}=0.2$ mm, such as in
Refs. \onlinecite{mchughetal07prb,hughetal09prb-tuning,hughetal09prb-species} is shown in
Fig. \ref{Fig-v_Bz=00003D0.7-1.6T-Myriam}.
In this case tunneling peaks are not washed out, although they are
much wider and lower than those in Fig. \ref{Fig-v_Estimation_BzExt=00003D0.7-1.5T}.
Some of these peaks are asymmetric, similarly to the single large
peak in Fig. 4 of Ref. \onlinecite{garjaa10prbrc}. The reason for this
asymmetry will be discussed below. Then, Fig. \ref{Fig-v_Bz=00003D1.5-3.5T}
shows the calculated front speed for the bias and crystal size corresponding
to the experiments in Refs. \onlinecite{mchughetal07prb,hughetal09prb-tuning,hughetal09prb-species}.
Here tunneling peaks are quite pronounced, at variance with the above
experiments that show very small peaks. Just above 3 T and just below
3.5 T there are regions where the speed is too high to be measured
in this calculation, an effect of ground-state tunneling.

\begin{figure}
\includegraphics[angle=-90,width=8cm]{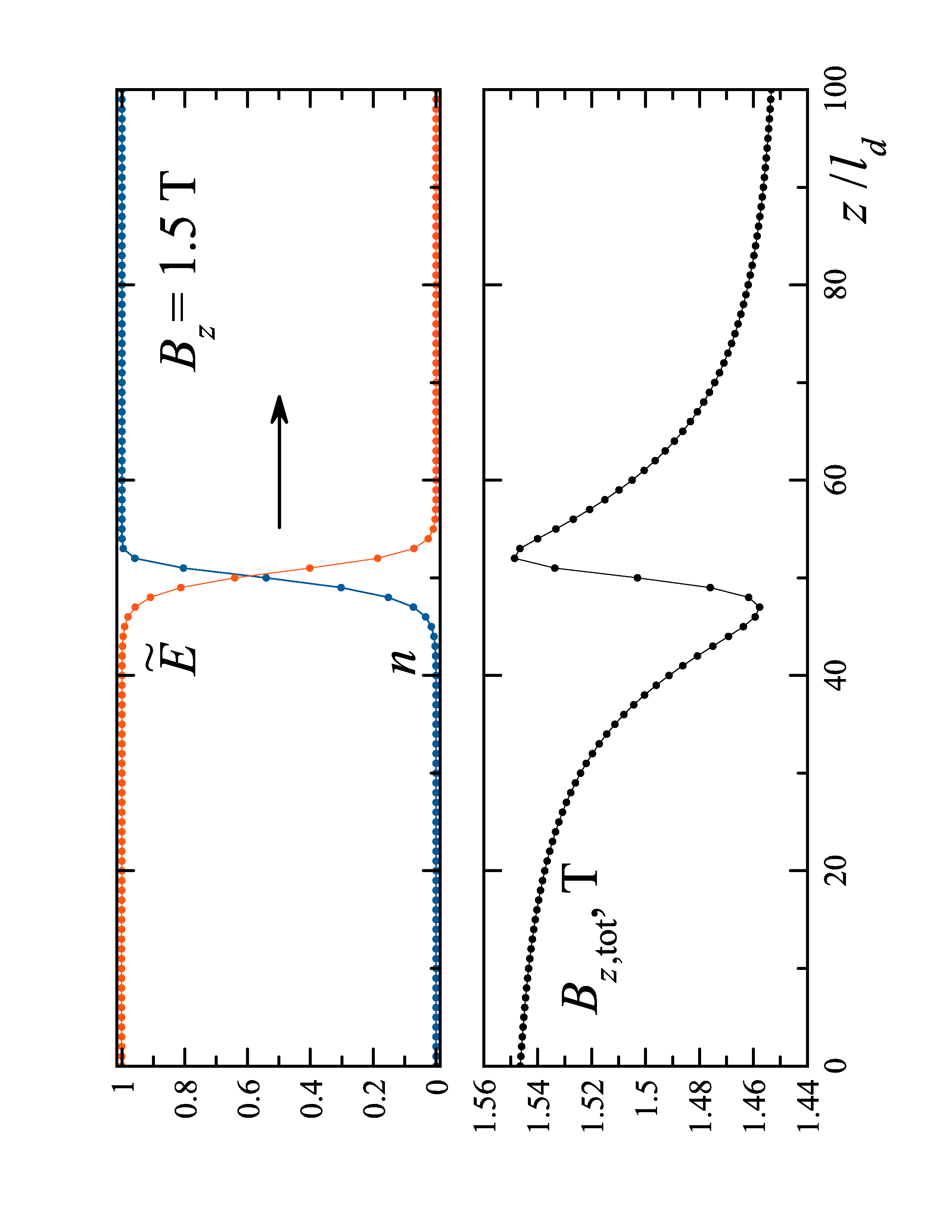}

\caption{Spatial profiles of the deflagration front in a small transverse field,
$B_{\bot}=0.05$ T. The bias $B_{z}=1.5$T is far from resonances,
thus the front is that of a pure slow burning.\label{Fig-Profile-Btr=00003D0.05T_BzExt=00003D1.5T}}

\end{figure}

Spatial profiles of the magnetization, energy, and the total bias
field in the deflagration front give an idea of the role played by
spin tunneling. Fig. \ref{Fig-Profile-Btr=00003D0.05T_BzExt=00003D1.5T} shows these profiles
at $B_{z}=1.5$ T that is far from resonances. In this case there
is a pure slow burning with the magnetization and energy profiles
of a tanh shape.\cite{garchu07prb} Dipolar field shown in the lower
panel plays no role in the process.

\begin{figure}

\includegraphics[angle=-90,width=8cm]{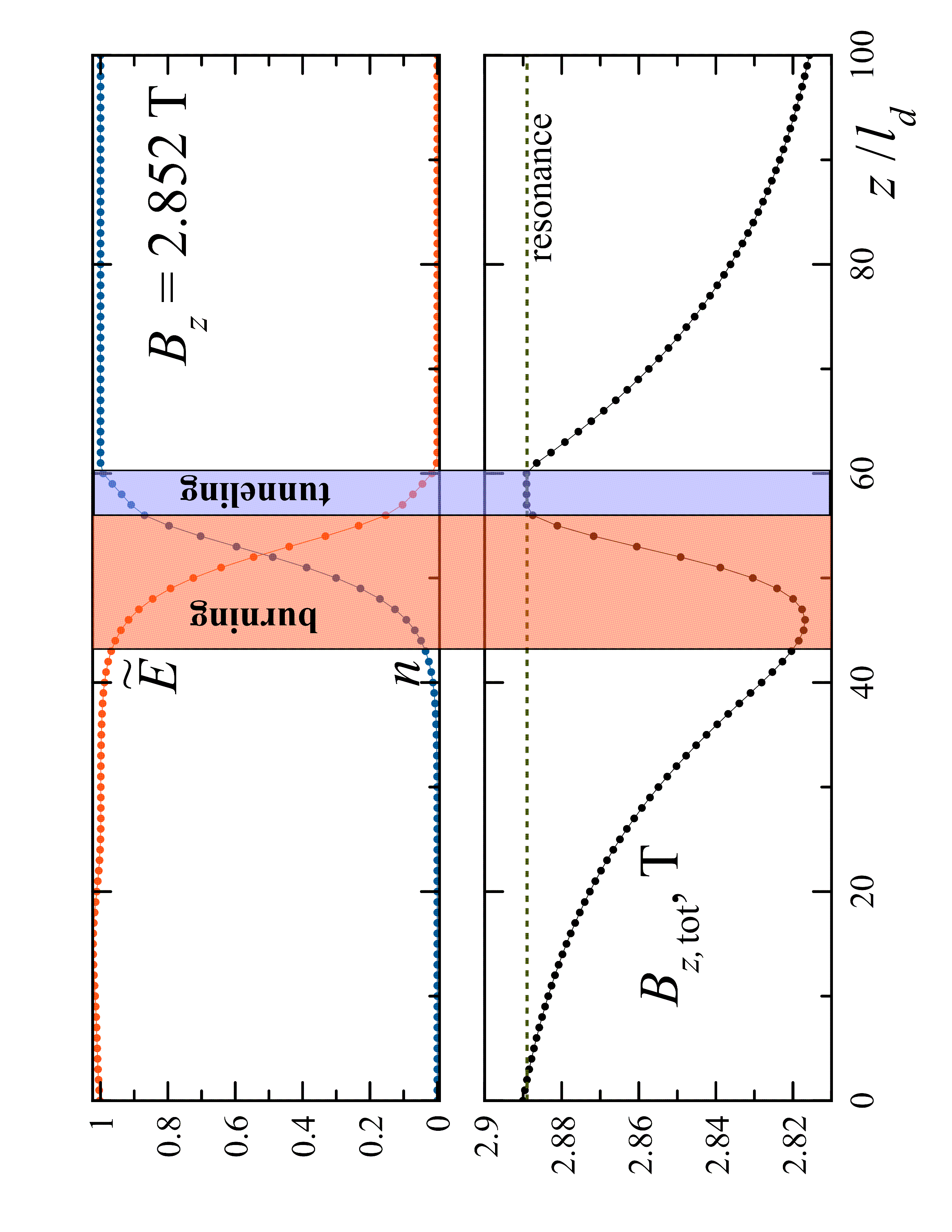}\caption{Spatial profiles of the deflagration front in a small transverse field,
$B_{\bot}=0.05$ T at the peak of the front speed at $B_{z}=2.852$
T. There is a resonance spin tunneling at the face of the front and
burning in its central and rear parts.}
\label{Fig-Profile-Btr=00003D0.05T_BzExt=00003D2.852T}

\end{figure}

Fig. \ref{Fig-Profile-Btr=00003D0.05T_BzExt=00003D2.852T} shows the
spatial profiles at the asymmetric peak of $v$ at $B_{z}=2.852$
T in Fig. \ref{Fig-v_Bz=00003D1.5-3.5T}. Here the front speed is
high because of tunneling at the face of the front where in the lower
panel the total bias field is flat at the level of the tunneling resonance
at $B_{z,\mathrm{tot}}=2.889$ T. Magnetization distribution adjusts
so that the dipolar field ensures resonance for a sizable group of
spins that tunnel. Tunneling of these spins results in energy release,
the temperature and relaxation rate increase, and tunneling gives
way to burning in the central and rear areas of the front.

Formation of the asymmetric maxima of the front speed can be explained
as follows. When $B_{z}$ increases, the peak of $B_{z,\mathrm{tot}}$
that arizes due to the local dipolar field reaches the resonant value.
Here the strong increase of $v(B_{z})$ begins. The maximum of $B_{z,\mathrm{tot}}$
sticks to the resonance value and becomes flat with progressively
increasing width. Greater width of the resonance region results in
a stronger tunneling and higher front speed. With further increase
of $B_{z}$, the right edge of the tunneling region moves too far
away from the front core into the region where the temperature is
too low. As the tunneling resonance in question is thermally assisted,
it disappears at low temperatures, thus the flat region of $B_{z,\mathrm{tot}}$
cannot spread too far to the right. As a result, the flat configuration
of $B_{z,\mathrm{tot}}$ becomes unstable and suddenly $B_{z,\mathrm{tot}}$
changes to the regular shape of Fig. \ref{Fig-Dzz-profile} that crosses
the resonance twice in the face part of the front. At the right crossing
the temperature is too low and tunneling does not occur, whereas at
the left crossing burning already is going on and tunneling cannot
add much. There can be the third resonance crossing further to the
left but it does not play a role because everything has already burned.
It should be noted that multiple resonance crossings do not occur
in the laminar regime of the pure quantum case (cold deflagration),
see Fig. 2 of Ref. \onlinecite{gar09prb}.

If the transverse size of the crystal is large, $R\gg l_{d}$, the slope
of $B_{z,\mathrm{tot}}$ to the right of the maximum in Fig. \ref{DzzCylinh}
is small. In this case increasing $B_{z}$ leads to a very quick displacement
of the right border of the tunneling region to the right where tunneling
cannot take place, as explained above. Thus tunneling peaks of $v(B_{z})$
should be very narrow for such crystals. This explains why tunneling
peaks are quite pronounced in Fig. \ref{Fig-v_Bz=00003D0.7-1.6T-Myriam}
but very small in Fig. \ref{Fig-v_Bz=00003D0.7-1.6T-Javie}.

Tunneling peaks of $v(B_{z})$ corresponding to broad resonances of
highly excited states are almost symmetric, such as the high peak
at $B_{z}=2.644$ T in Fig. \ref{Fig-v_Bz=00003D1.5-3.5T}. In this
case peaks are formed when the maximum of $B_{z,\mathrm{tot}}$ crosses
the resonance. In these cases progressive flattening of $B_{z,\mathrm{tot}}$
does not occur because here tunneling requires high temperatures and
the right border of the tunneling region cannot move to the cold region
to the right.

\begin{figure}
\includegraphics[angle=-90,width=8cm]{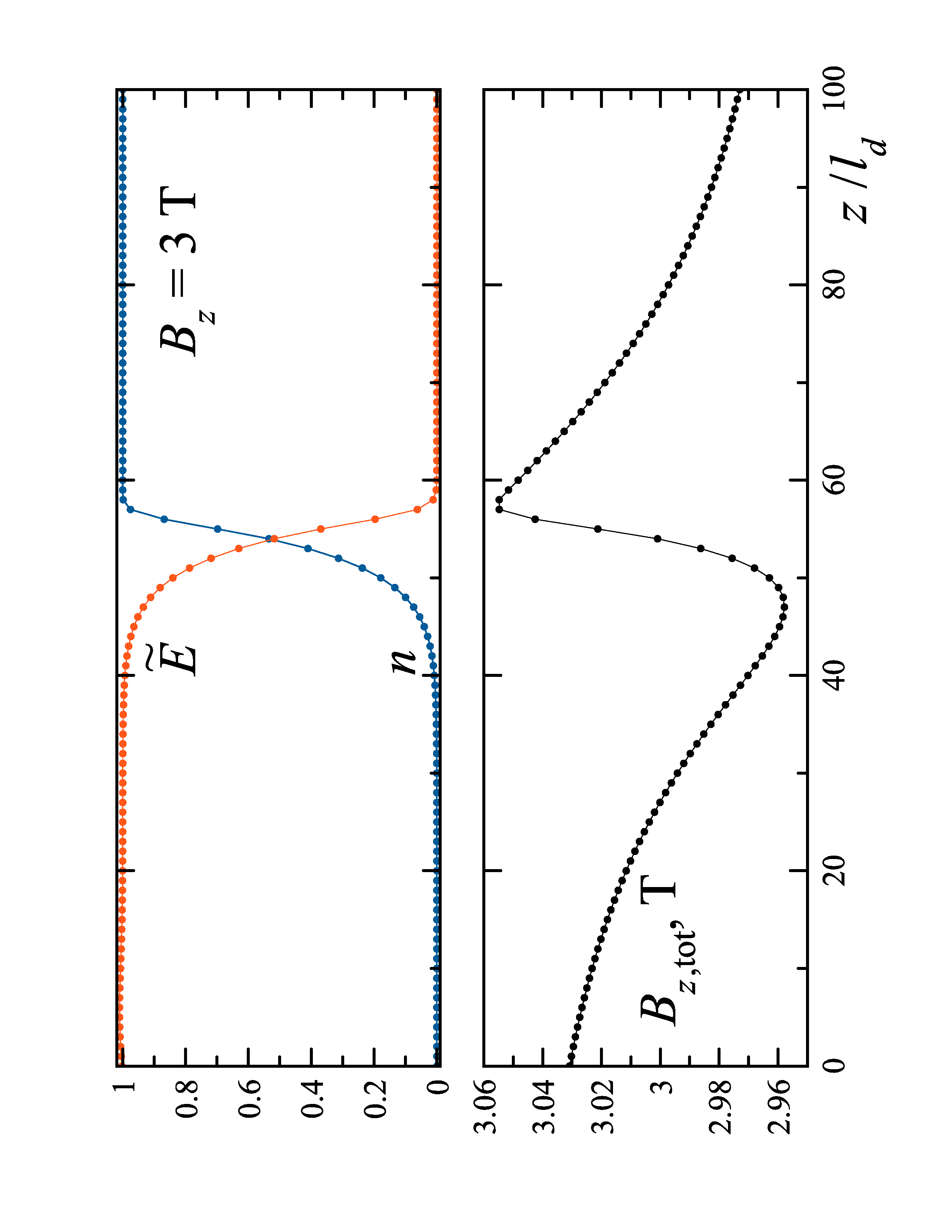}

\caption{Spatial profiles of the deflagration front in a small transverse field,
$B_{\bot}=0.05$ T. Faster burning with no tunneling.}
\label{Fig-Profile-Btr=00003D0.05T_BzExt=00003D3T}

\end{figure}

Fig. \ref{Fig-Profile-Btr=00003D0.05T_BzExt=00003D3T} shows an off-resonance
front again, $B_{z}=3$ T, but it is not a slow burning front anymore
because the bias is high. In this region the analytical theory of
Ref. \onlinecite{garchu07prb} does not work that can be seen on the magnetization
and energy profiles that differ from the tanh shape.

\section{Front speed at strong transverse field\label{sec:Front-speed-strong-field}}

As one can see from Fig. \ref{Fig-3D-Gamma_Btr=00003D3.5T}, at strong
transverse fields the structure of the relaxation rate $\Gamma(B_{z},T)$
simplifies because tunneling resonances of the most excited states
broaden away. At $B_{\bot}=3.5$ T one can see only two tunneling
peaks, and the ground-state tunneling peak is not drowned by the thermal-activation
processes up to the highest temperatures. This means that in a bias
window around this peak, the barrier is cut completely. The latter
changes the dynamics of the system, drastically increasing the role
of tunneling in the front propagation. Since tunneling out of the
metastable ground state does not require an elevated temperature,
the right border of the tunneling region before the main part of the
front can shift unlimitedly to the right without causing the instability
that kills tunneling, described in the preceding section. Thus the
width of the tunneling region can reach the values of order $R$,
\cite{garchu09prl,gar09prb} that leads to front speeds much greater
than the speed of a regular magnetic deflagration {[}see comments
after Eq. (\ref{vstar}){]}.

\begin{figure}
,\includegraphics[angle=-90,width=8cm]{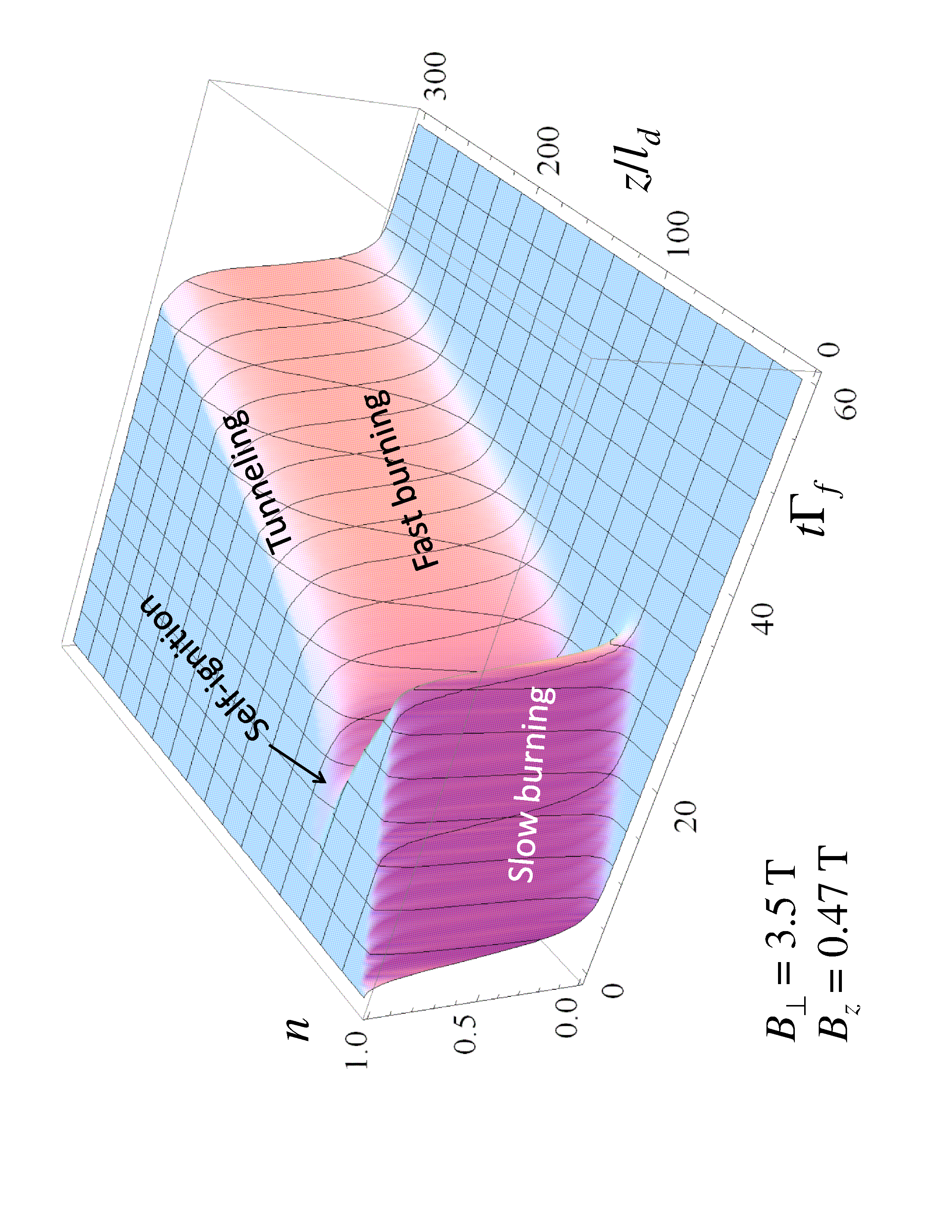}

\caption{Quantum self-ignition before a slow-burning front, leading to a fast combined
tunneling-burning front near ground-state resonance at high transverse
fields.}

\label{Fig-3D-Front-slow+fast}
\end{figure}

\begin{figure}
\includegraphics[angle=-90,width=8cm]{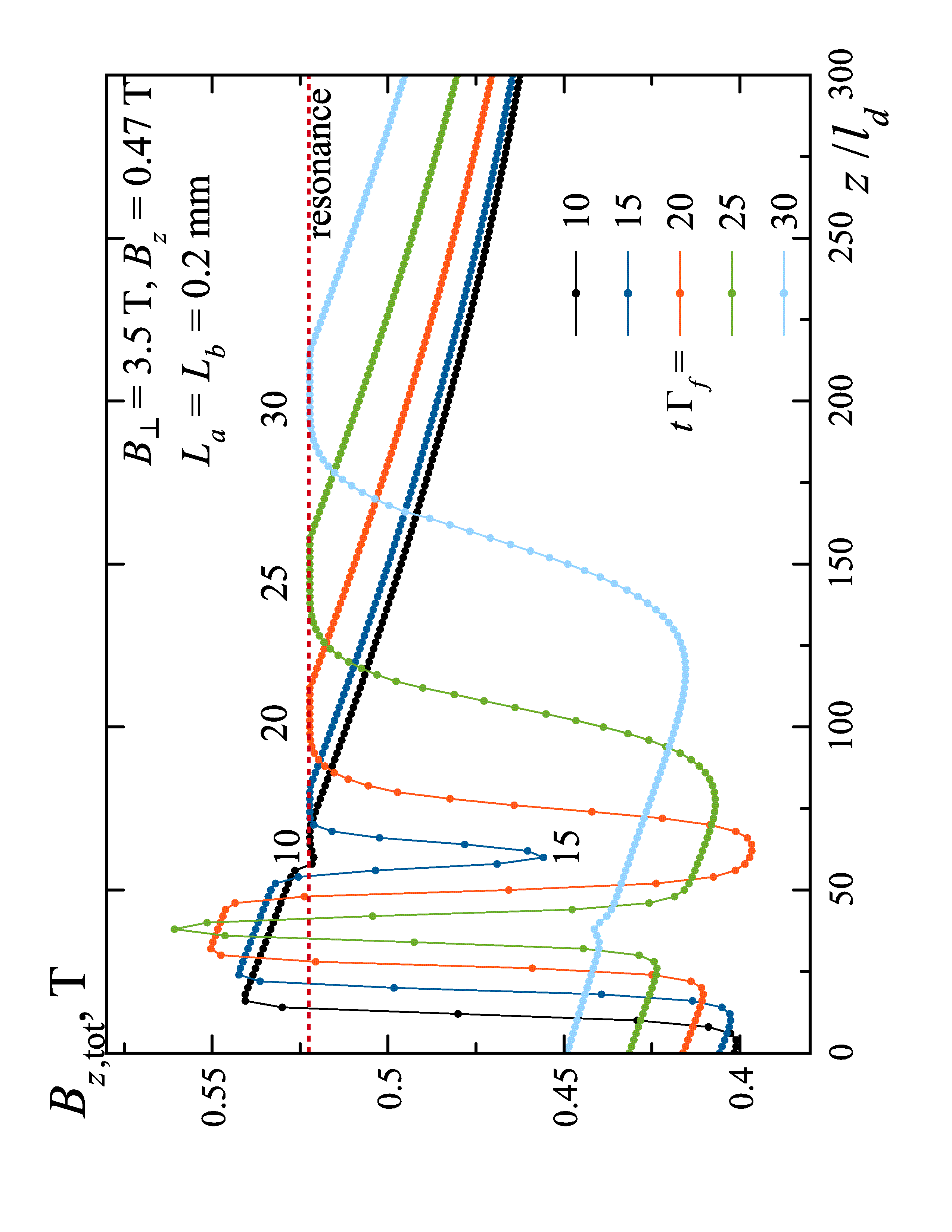}

\caption{Profiles of the total longitudinal magnetic field at $B_{\bot}=3.5$
T and $B_{z}=0.47$ T at different moments of time.}
\label{Fig-Profiles_Bz_Btr=00003D3.5T_BzExt=00003D0.47T_Ltil=00003D300_N=00003D150}

\end{figure}

Numerical results at high transverse fields show that shortly after
ignition by raising the temperature at the left end of the crystal,
a regular slow-burning front can transform into a fast combined tunneling-burning
front by \emph{quantum self-ignition} before the slow-burning front,
if the crystal is near ground-state tunneling resonance. Fig. \ref{Fig-3D-Front-slow+fast}
shows this phenomenon at $B_{\bot}=3.5$ T and $B_{z}=0.47$ T, where
the ground-state resonance is achieved at $B_{z,\mathrm{tot}}=0.522$
T. One can see that at short times, $t\Gamma_{f}\simeq10$, there
is a slow front with a steep profile but before the front, where $B_{z,\mathrm{tot}}$
crosses the resonance value, spins begin to tunnel. This quantum self-ignition
leads to flattening of the $B_{z,\mathrm{tot}}$ curve and formation
of another, fast moving front, with tunneling followed by burning.
The spatial profile of $B_{z,\mathrm{tot}}$ at different times that
shows that self-ignition before the slow-burning front is caused by
spin tunneling, is shown in Fig. \ref{Fig-Profiles_Bz_Btr=00003D3.5T_BzExt=00003D0.47T_Ltil=00003D300_N=00003D150}.

\begin{figure}
\includegraphics[angle=-90,width=8cm]{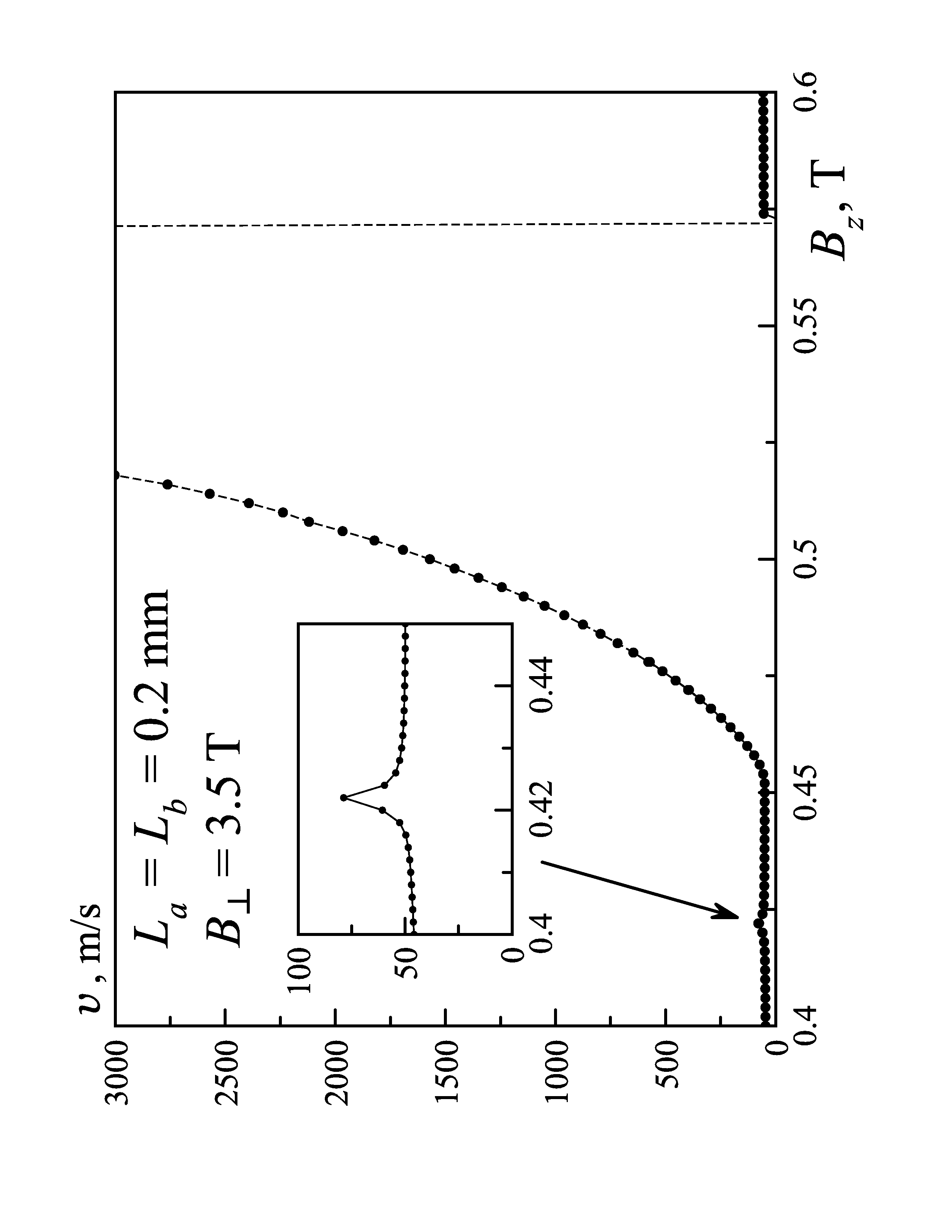}

\caption{Front speed for a strong transverse field ($B_{\bot}=3.5$ T) in the
vicinity of the ground-state tunneling resonance at 0.522 T.
There is a strong increase of the front speed within the dipolar window of 125.5 mT around the resonance.
The small
peak on the left (inset) is due to the first-excited-state tunneling
resonance. }
\label{Fig-v_Btr=00003D3.5T_Bz=00003D0.4-0.6T}

\end{figure}

The front speed $v$ in the vicinity of a biased ground-state resonance
in a strong transverse field can achieve supersonic values, as can
be seen in Fig. \ref{Fig-v_Btr=00003D3.5T_Bz=00003D0.4-0.6T}. This
is in accord with the comments below Eq. (\ref{vstar}). To the contrast,
the small peak on the left side of Fig. \ref{Fig-v_Btr=00003D3.5T_Bz=00003D0.4-0.6T}
is due to the first-excited-state tunneling resonance at $B_{k}=0.490$,
see Fig. \ref{Fig-3D-Gamma_Btr=00003D3.5T}. Its position is given
by $B_{z}=B_{k}-B_{z}^{(k_{D})}=0.417$ T that is close to the position
in the figure. In Sec. \ref{sec:Front-speed-weak-field} it was explaind
that front of tunneling via excited levels cannot shift much ahead
of the burning zone because it is too cold before the front. This
limits the speed of such fronts and explains why the speed of the
first-excited state tunneling front is much smaller than that of the
ground-state tunneling front, in spite of the relaxation rate at the
former being higher.

Returning to the ground-state tunneling front, it should be stressed
that no metastable population is left behind the front (see Fig. \ref{Fig-3D-Front-slow+fast}),
although there is unburned metastable population behind pure non-thermal
fronts of tunneling.\cite{garchu09prl,gar09prb} Here, the metastable
population is burning just behind the front of tunneling as the result
of the temperature increase. It should be stressed that heat conduction
cannot support burning fronts moving faster than the speed of sound
 and it becomes non-operative in this case. In this respect,
the situation is reminding of detonation that has been suggested for
molecular magnets in Refs. \onlinecite{modbycmar11prb,modbycmar11prl}
in the case of a strong bias and thus high energy release. In detonation,
thermal expansion resulting from burning sends a shock wave into the
cold region before the front where, as a consequence, the temperature
rises as a result of compression, initiating burning. As the mechanism
of detonation is based on elasticity, the speed of a detonation
front is comparable to the speed of sound. Fronts of tunneling are
not based on elasticity and their speed can be much higher. However,
shock waves must accompany tunneling fronts and modify their properties
in some way. Experimentally, fast deflagration or detonation fronts
in Mn$_{12}$ Ac have been observed in Ref. \onlinecite{decvanmostejhermac09prl}
but they were caused by a very fast sweep of $B_{z}$, so that there
is a question to which extent the process was self-propelled.

One can see in Fig. \ref{Fig-v_Btr=00003D3.5T_Bz=00003D0.4-0.6T}
that the speed of the front is asymmetric and grows toward the right
end of the tunneling window, showing divergence or nearly divergence
of the front speed. In the case of cold deflagration (assuming the
unbroken laminar regime everywhere) $v$ diverges at the right border
of the dipolar window, $B_{z}=B_{k}+B_{z}^{(D)}$, where $B_{k}$
is the resonance field of Eq. (\ref{Bk}). It is given by \cite{gar09prb}
\begin{equation}
v\sim\Gamma_{\mathrm{res}}R\frac{B_{z}-B_{k}}{B_{k}+B_{z}^{(D)}-B_{z}},\label{vDiverging}
\end{equation}
for $B_{k}\leq B_{z}\leq B_{k}+B_{z}^{(D)}$, whereas above $B_{k}+B_{z}^{(D)}$
it abruptly drops to a zero.
The reason for this is that above $B_{k}+B_{z}^{(D)}$,
the total field well before the front is above the resonance, so that
resonance crossing cannot occur. To the contrast, just below $B_{k}+B_{z}^{(D)}$
the field well before the front is a little bit below the resonance
and increases closer to the front. In this case, there is a wide region
where the system is close to the resonance, and the front speed becomes
very high. Thus as $B_{z}$ crosses the value $B_{k}+B_{z}^{(D)}$ from below, the
front speed drops abruptly. Similar
behavior can be seen in Fig. \ref{Fig-v_Btr=00003D3.5T_Bz=00003D0.4-0.6T}:
The ground-state resonance is at 0.522 T, and adding the dipolar field
$B_{z}^{(D)}=0.0526$ T one obtains 0.573 T, as in the figure. However,
here $v$ drops to the speed of the regular magnetic deflagration,
as also in Fig. 4 of Ref. \onlinecite{garjaa10prbrc}. Another difference
is that in the case of cold deflagration\cite{gar09prb} tunneling begins at $B_{z}=B_{k}$
(left border of the dipolar window) whereas in our case it begins
when the local maximum of $B_{z\mathrm{,tot}}$ first touches the
resonance (see, e.g., the lower panel of Fig. \ref{Fig-Profile-Btr=00003D0.05T_BzExt=00003D2.852T}).
Since for the crystals studied here the front width is much smaller than
the transverse size of the crystal, the dipolar field at the maximum
is close to $B_{z}^{(k_{D})}$ given by Eq. (\ref{BzkD}). Thus the
left border of the dipolar window is at $B_{z}=B_{k}-B_{z}^{(k_{D})}$,
in Fig. \ref{Fig-v_Btr=00003D3.5T_Bz=00003D0.4-0.6T} at $B_{z}=0.45$
T. The total width of the dipolar window of the ground-state tunneling
resonance in Mn$_{12}$ Ac is
\begin{equation}
\Delta B_{z}^{(D)}=B_{z}^{(D)}+B_{z}^{(k_{D})}=125.5\,\mathrm{mT}\label{DeltaBzD}
\end{equation}
 that is much greater than dipolar windows of excited-state tunneling
resonances, see, e.g., Fig. \ref{Fig-v_Bz=00003D1.5-3.5T}.

In the case of cold deflagration, there is an unburned metastable
population in the final state behind the front, Eq. (41) of Ref. \onlinecite{gar09prb}
that can be rewritten as
\begin{equation}
n_{f}=\frac{B_{z}-B{}_{k}}{B_{z}^{(D)}}\label{nf}
\end{equation}
($n=1$ before the front). One can see that the change of $n$ across
the front $\Delta n=1-n_{f}$ goes to zero at the right border of
the dipolar window $B_{k}+B_{z}^{(D)}$. This reconciles the situation
with the general requirement that the rate of change of the magnetization
of the crystal $\dot{M}$, limited by the tunneling parameter $\Delta$,
remains finite. Indeed,
\begin{equation}
\dot{M}\propto(1-n_{f})v=\Gamma_{\mathrm{res}}R\frac{B_{z}-B_{k}}{B_{z}^{(D)}}\label{Mdot}
\end{equation}
reaches only a finite value $\dot{M}\propto\Gamma_{\mathrm{res}}R$
at the right border of the dipolar window before it drops to zero.
In the present case of tunneling followed by complete burning, $\dot{M}$
is not limited by $\Delta$ and can achieve very high values at the
right border of the dipolar window.

\section{Discussion\label{sec:Discussion}}

Numerical calculations for deflagration fronts with dipolar-controlled
spin tunneling for the realistic model of Mn$_{12}$ Ac performed
in this work have shown many quantum peaks in the dependence of the
front speed $v$ on the external magnetic field $B_{z}$, if a zero
or small transverse field is applied. The multitude of peaks results
from the splitting of the tunneling resonance by the $-AS_{z}^{4}$
term in the crystal field of the magnetic molecule, and peaks in $v(B_{z})$
reflect those in the relaxation rate $\Gamma(B_{z},T)$ of the metastable
states of Mn$_{12}$ Ac molecules. The peaks of $v(B_{z})$ are more
pronounced for crystals of a smaller transverse size.

Whereas the results of the calculations for thicker crystals in the
range of smaller bias are in a qualitative accord with the experiments
of Ref. \onlinecite{heretal05prl}, the results for thinner crystals and
stronger bias show much stronger tunneling peaks in $v(B_{z})$ than
it was observed in Refs. \onlinecite{mchughetal07prb,hughetal09prb-tuning,hughetal09prb-species}.
One can try to explain the lack of peaks in the experiment by the
spread of tunneling resonances as the result of ligand disorder that
is pretty strong in Mn$_{12}$ Ac.\cite{parketal01prb,parketal02prb}
It has been shown that static disorder that is weaker than the dipole-dipole
interaction does not destroy fronts of tunneling since the magnetization
distribution can adjust so that many spins in the front core are still
on resonance and can tunnel.\cite{garchu09prl} However, static disorder
that is stronger than the DDI cannot be accomodated by the latter
and should result in spread and suppression of tunneling maxima in
$v(B_{z})$. The best way to deal with this problem is to make experiments
on the members of Mn$_{12}$ family that do not have ligand disorder.

Possibly there is a more fundamental reason for the near absence of
tunneling peaks in the experiments of Refs. \onlinecite{mchughetal07prb,hughetal09prb-tuning,hughetal09prb-species}.
The prefactor $\Gamma_{0}$ in the theoretical relaxation rate being
by a factor 10$^{2}$ smaller than the measured prefactor suggests
collective relaxation processes such as superradiance and phonon bottleneck
that can be expected in a dense magnetic system such as molecular
magnet. Collective boosting of relaxation processes should not affect
tunneling, however. Thus the non-resonant background in $\Gamma(B_{z},T)$
can move up by a factor 100, partially drowning tunneling peaks. On
the other hand, in our calculation we have boosted the whole function
$\Gamma(B_{z},T)$, including tunneling peaks.

Whereas it is impossible to develop a collective theory of relaxation
in molecular magnets in this paper, one can modify our density-matrix
calculation by multiplying spin-phonon coupling amplitudes by 10.
This should result in the increase of the non-resonant part of $\Gamma(B_{z},T)$
by 100, while one could expect tunneling peaks to be much less changed.
Such a calculation has been performed but its results do not
show a strong suppression of tunneling peaks in $v(B_{z})$. The likely
reason for this is that at small transverse field tunneling peaks
are due to thermally assisted tunneling that also gets boosted by
an artificial increase of spin-phonon interactions.

Calculations in the case of a strong transverse field, making
tunneling directly out of the metastable ground state operative, show
an increase of the front speed within the tunneling window around
the tunneling resonance up to supersonic values. It would be highly interesting to perform
experiments on deflagration fronts in this region.

This work has been supported by the NSF under Grant No. DMR-0703639.
The author thanks E. M. Chudnovsky for valuable discussions.

\bibliography{chu-own,gar-own,gar-relaxation,gar-tunneling,gar-MM-ordering,gar-books,gar-oldworks}

\begin{thebibliography}{47}
\expandafter\ifx\csname natexlab\endcsname\relax\def\natexlab#1{#1}\fi
\expandafter\ifx\csname bibnamefont\endcsname\relax
  \def\bibnamefont#1{#1}\fi
\expandafter\ifx\csname bibfnamefont\endcsname\relax
  \def\bibfnamefont#1{#1}\fi
\expandafter\ifx\csname citenamefont\endcsname\relax
  \def\citenamefont#1{#1}\fi
\expandafter\ifx\csname url\endcsname\relax
  \def\url#1{\texttt{#1}}\fi
\expandafter\ifx\csname urlprefix\endcsname\relax\def\urlprefix{URL }\fi
\providecommand{\bibinfo}[2]{#2}
\providecommand{\eprint}[2][]{\url{#2}}

\bibitem[{\citenamefont{Glassman}(1996)}]{gla96book}
\bibinfo{author}{\bibfnamefont{I.}~\bibnamefont{Glassman}},
  \emph{\bibinfo{title}{Combustion}} (\bibinfo{publisher}{Academic Press},
  \bibinfo{year}{1996}).

\bibitem[{\citenamefont{Landau and Lifshitz}(1987)}]{lanlif9fluid}
\bibinfo{author}{\bibfnamefont{L.~D.} \bibnamefont{Landau}} \bibnamefont{and}
  \bibinfo{author}{\bibfnamefont{E.~M.} \bibnamefont{Lifshitz}},
  \emph{\bibinfo{title}{Fluid {D}ynamics}} (\bibinfo{publisher}{Pergamon},
  \bibinfo{address}{London}, \bibinfo{year}{1987}).

\bibitem[{\citenamefont{Suzuki et~al.}(2005)\citenamefont{Suzuki, Sarachik,
  Chudnovsky, McHugh, Gonzalez-Rubio, Avraham, Myasoedov, Zeldov, Shtrikman,
  Chakov et~al.}}]{suzetal05prl}
\bibinfo{author}{\bibfnamefont{Y.}~\bibnamefont{Suzuki}},
  \bibinfo{author}{\bibfnamefont{M.~P.} \bibnamefont{Sarachik}},
  \bibinfo{author}{\bibfnamefont{E.~M.} \bibnamefont{Chudnovsky}},
  \bibinfo{author}{\bibfnamefont{S.}~\bibnamefont{McHugh}},
  \bibinfo{author}{\bibfnamefont{R.}~\bibnamefont{Gonzalez-Rubio}},
  \bibinfo{author}{\bibfnamefont{N.}~\bibnamefont{Avraham}},
  \bibinfo{author}{\bibfnamefont{Y.}~\bibnamefont{Myasoedov}},
  \bibinfo{author}{\bibfnamefont{E.}~\bibnamefont{Zeldov}},
  \bibinfo{author}{\bibfnamefont{H.}~\bibnamefont{Shtrikman}},
  \bibinfo{author}{\bibfnamefont{N.~E.} \bibnamefont{Chakov}},
  \bibnamefont{et~al.}, \bibinfo{journal}{Phys. Rev. Lett.}
  \textbf{\bibinfo{volume}{95}}, \bibinfo{pages}{147201}
  (\bibinfo{year}{2005}).

\bibitem[{\citenamefont{Hern\'andez-Minguez
  et~al.}(2005)\citenamefont{Hern\'andez-Minguez, Hern\'andez, Macia,
  Garcia-Santiago, Tejada, and Santos}}]{heretal05prl}
\bibinfo{author}{\bibfnamefont{A.}~\bibnamefont{Hern\'andez-Minguez}},
  \bibinfo{author}{\bibfnamefont{J.~M.} \bibnamefont{Hern\'andez}},
  \bibinfo{author}{\bibfnamefont{F.}~\bibnamefont{Macia}},
  \bibinfo{author}{\bibfnamefont{A.}~\bibnamefont{Garcia-Santiago}},
  \bibinfo{author}{\bibfnamefont{J.}~\bibnamefont{Tejada}}, \bibnamefont{and}
  \bibinfo{author}{\bibfnamefont{P.~V.} \bibnamefont{Santos}},
  \bibinfo{journal}{Phys. Rev. Lett.} \textbf{\bibinfo{volume}{95}},
  \bibinfo{pages}{217205} (\bibinfo{year}{2005}).

\bibitem[{\citenamefont{Garanin and Chudnovsky}(2007)}]{garchu07prb}
\bibinfo{author}{\bibfnamefont{D.~A.} \bibnamefont{Garanin}} \bibnamefont{and}
  \bibinfo{author}{\bibfnamefont{E.~M.} \bibnamefont{Chudnovsky}},
  \bibinfo{journal}{Phys. Rev. B} \textbf{\bibinfo{volume}{76}},
  \bibinfo{pages}{054410} (\bibinfo{year}{2007}).

\bibitem[{\citenamefont{McHugh et~al.}(2007)\citenamefont{McHugh, Jaafar,
  Sarachik, Myasoedov, Finkler, Shtrikman, Zeldov, Bagai, and
  Christou}}]{mchughetal07prb}
\bibinfo{author}{\bibfnamefont{S.}~\bibnamefont{McHugh}},
  \bibinfo{author}{\bibfnamefont{R.}~\bibnamefont{Jaafar}},
  \bibinfo{author}{\bibfnamefont{M.~P.} \bibnamefont{Sarachik}},
  \bibinfo{author}{\bibfnamefont{Y.}~\bibnamefont{Myasoedov}},
  \bibinfo{author}{\bibfnamefont{A.}~\bibnamefont{Finkler}},
  \bibinfo{author}{\bibfnamefont{H.}~\bibnamefont{Shtrikman}},
  \bibinfo{author}{\bibfnamefont{E.}~\bibnamefont{Zeldov}},
  \bibinfo{author}{\bibfnamefont{R.}~\bibnamefont{Bagai}}, \bibnamefont{and}
  \bibinfo{author}{\bibfnamefont{G.}~\bibnamefont{Christou}},
  \bibinfo{journal}{Phys. Rev. B} \textbf{\bibinfo{volume}{76}},
  \bibinfo{pages}{172410} (\bibinfo{year}{2007}).

\bibitem[{\citenamefont{McHugh et~al.}(2009{\natexlab{a}})\citenamefont{McHugh,
  Wen, Ma, Sarachik, Myasoedov, Zeldov, Bagai, and
  Christou}}]{hughetal09prb-tuning}
\bibinfo{author}{\bibfnamefont{S.}~\bibnamefont{McHugh}},
  \bibinfo{author}{\bibfnamefont{B.}~\bibnamefont{Wen}},
  \bibinfo{author}{\bibfnamefont{X.}~\bibnamefont{Ma}},
  \bibinfo{author}{\bibfnamefont{M.~P.} \bibnamefont{Sarachik}},
  \bibinfo{author}{\bibfnamefont{Y.}~\bibnamefont{Myasoedov}},
  \bibinfo{author}{\bibfnamefont{E.}~\bibnamefont{Zeldov}},
  \bibinfo{author}{\bibfnamefont{R.}~\bibnamefont{Bagai}}, \bibnamefont{and}
  \bibinfo{author}{\bibfnamefont{G.}~\bibnamefont{Christou}},
  \bibinfo{journal}{Phys. Rev. B} \textbf{\bibinfo{volume}{79}},
  \bibinfo{pages}{174413} (\bibinfo{year}{2009}{\natexlab{a}}).

\bibitem[{\citenamefont{McHugh et~al.}(2009{\natexlab{b}})\citenamefont{McHugh,
  Jaafar, Sarachik, Myasoedov, Finkler, Zeldov, Bagai, and
  Christou}}]{hughetal09prb-species}
\bibinfo{author}{\bibfnamefont{S.}~\bibnamefont{McHugh}},
  \bibinfo{author}{\bibfnamefont{R.}~\bibnamefont{Jaafar}},
  \bibinfo{author}{\bibfnamefont{M.~P.} \bibnamefont{Sarachik}},
  \bibinfo{author}{\bibfnamefont{Y.}~\bibnamefont{Myasoedov}},
  \bibinfo{author}{\bibfnamefont{A.}~\bibnamefont{Finkler}},
  \bibinfo{author}{\bibfnamefont{E.}~\bibnamefont{Zeldov}},
  \bibinfo{author}{\bibfnamefont{R.}~\bibnamefont{Bagai}}, \bibnamefont{and}
  \bibinfo{author}{\bibfnamefont{G.}~\bibnamefont{Christou}},
  \bibinfo{journal}{Phys. Rev. B} \textbf{\bibinfo{volume}{80}},
  \bibinfo{pages}{024403} (\bibinfo{year}{2009}{\natexlab{b}}).

\bibitem[{\citenamefont{Maci\`a et~al.}(2007)\citenamefont{Maci\`a,
  Hern\'andez-M\'\i{}nguez, Abril, Hernandez, Garc\'\i{}a-Santiago, Tejada,
  Parisi, and Santos}}]{masiaetal07prb}
\bibinfo{author}{\bibfnamefont{F.}~\bibnamefont{Maci\`a}},
  \bibinfo{author}{\bibfnamefont{A.}~\bibnamefont{Hern\'andez-M\'\i{}nguez}},
  \bibinfo{author}{\bibfnamefont{G.}~\bibnamefont{Abril}},
  \bibinfo{author}{\bibfnamefont{J.~M.} \bibnamefont{Hernandez}},
  \bibinfo{author}{\bibfnamefont{A.}~\bibnamefont{Garc\'\i{}a-Santiago}},
  \bibinfo{author}{\bibfnamefont{J.}~\bibnamefont{Tejada}},
  \bibinfo{author}{\bibfnamefont{F.}~\bibnamefont{Parisi}}, \bibnamefont{and}
  \bibinfo{author}{\bibfnamefont{P.~V.} \bibnamefont{Santos}},
  \bibinfo{journal}{Phys. Rev. B} \textbf{\bibinfo{volume}{76}},
  \bibinfo{pages}{174424} (\bibinfo{year}{2007}).

\bibitem[{\citenamefont{Decelle et~al.}(2009)\citenamefont{Decelle, Vanacken,
  Moshchalkov, Tejada, Hern\'andez, and Maci\`a}}]{decvanmostejhermac09prl}
\bibinfo{author}{\bibfnamefont{W.}~\bibnamefont{Decelle}},
  \bibinfo{author}{\bibfnamefont{J.}~\bibnamefont{Vanacken}},
  \bibinfo{author}{\bibfnamefont{V.~V.} \bibnamefont{Moshchalkov}},
  \bibinfo{author}{\bibfnamefont{J.}~\bibnamefont{Tejada}},
  \bibinfo{author}{\bibfnamefont{J.~M.} \bibnamefont{Hern\'andez}},
  \bibnamefont{and} \bibinfo{author}{\bibfnamefont{F.}~\bibnamefont{Maci\`a}},
  \bibinfo{journal}{Phys. Rev. Lett.} \textbf{\bibinfo{volume}{102}},
  \bibinfo{pages}{027203} (\bibinfo{year}{2009}).

\bibitem[{\citenamefont{Modestov
  et~al.}(2011{\natexlab{a}})\citenamefont{Modestov, Bychkov, and
  Marklund}}]{modbycmar11prb}
\bibinfo{author}{\bibfnamefont{M.}~\bibnamefont{Modestov}},
  \bibinfo{author}{\bibfnamefont{V.}~\bibnamefont{Bychkov}}, \bibnamefont{and}
  \bibinfo{author}{\bibfnamefont{M.}~\bibnamefont{Marklund}},
  \bibinfo{journal}{Phys. Rev. B} \textbf{\bibinfo{volume}{83}},
  \bibinfo{pages}{214417} (\bibinfo{year}{2011}{\natexlab{a}}).

\bibitem[{\citenamefont{Modestov
  et~al.}(2011{\natexlab{b}})\citenamefont{Modestov, Bychkov, and
  Marklund}}]{modbycmar11prl}
\bibinfo{author}{\bibfnamefont{M.}~\bibnamefont{Modestov}},
  \bibinfo{author}{\bibfnamefont{V.}~\bibnamefont{Bychkov}}, \bibnamefont{and}
  \bibinfo{author}{\bibfnamefont{M.}~\bibnamefont{Marklund}},
  \bibinfo{journal}{Phys. Rev. Lett.} \textbf{\bibinfo{volume}{107}},
  \bibinfo{pages}{20720} (\bibinfo{year}{2011}{\natexlab{b}}).

\bibitem[{\citenamefont{{R. Sessoli, D. Gatteschi, A. Caneschi, and M. A.
  Novak}}(1993)}]{sesgatcannov93nat}
\bibinfo{author}{\bibnamefont{{R. Sessoli, D. Gatteschi, A. Caneschi, and M. A.
  Novak}}}, \bibinfo{journal}{Nature (London)} \textbf{\bibinfo{volume}{365}},
  \bibinfo{pages}{141} (\bibinfo{year}{1993}).

\bibitem[{\citenamefont{Barra et~al.}(1996)\citenamefont{Barra, Debrunner,
  Gatteschi, Schultz, and Sessoli}}]{baretal96epl}
\bibinfo{author}{\bibfnamefont{A.~L.} \bibnamefont{Barra}},
  \bibinfo{author}{\bibfnamefont{P.}~\bibnamefont{Debrunner}},
  \bibinfo{author}{\bibfnamefont{D.}~\bibnamefont{Gatteschi}},
  \bibinfo{author}{\bibfnamefont{C.~E.} \bibnamefont{Schultz}},
  \bibnamefont{and} \bibinfo{author}{\bibfnamefont{R.}~\bibnamefont{Sessoli}},
  \bibinfo{journal}{Europhys. Lett.} \textbf{\bibinfo{volume}{35}},
  \bibinfo{pages}{133} (\bibinfo{year}{1996}).

\bibitem[{\citenamefont{Gatteschi et~al.}(2006)\citenamefont{Gatteschi,
  Sessoli, and Villain}}]{gatsesvil06book}
\bibinfo{author}{\bibfnamefont{D.}~\bibnamefont{Gatteschi}},
  \bibinfo{author}{\bibfnamefont{R.}~\bibnamefont{Sessoli}}, \bibnamefont{and}
  \bibinfo{author}{\bibfnamefont{J.}~\bibnamefont{Villain}},
  \emph{\bibinfo{title}{Molecular {N}anomagnets}} (\bibinfo{publisher}{Oxford
  University Press}, \bibinfo{address}{Oxford}, \bibinfo{year}{2006}).

\bibitem[{\citenamefont{Friedman et~al.}(1996)\citenamefont{Friedman, Sarachik,
  Tejada, and Ziolo}}]{frisartejzio96prl}
\bibinfo{author}{\bibfnamefont{J.~R.} \bibnamefont{Friedman}},
  \bibinfo{author}{\bibfnamefont{M.~P.} \bibnamefont{Sarachik}},
  \bibinfo{author}{\bibfnamefont{J.}~\bibnamefont{Tejada}}, \bibnamefont{and}
  \bibinfo{author}{\bibfnamefont{R.}~\bibnamefont{Ziolo}},
  \bibinfo{journal}{Phys. Rev. Lett.} \textbf{\bibinfo{volume}{76}},
  \bibinfo{pages}{3830} (\bibinfo{year}{1996}).

\bibitem[{\citenamefont{Hern\'andez et~al.}(1996)\citenamefont{Hern\'andez,
  Zhang, Luis, Bartolom\'e, Tejada, and Ziolo}}]{heretal96epl}
\bibinfo{author}{\bibfnamefont{J.~M.} \bibnamefont{Hern\'andez}},
  \bibinfo{author}{\bibfnamefont{X.~X.} \bibnamefont{Zhang}},
  \bibinfo{author}{\bibfnamefont{F.}~\bibnamefont{Luis}},
  \bibinfo{author}{\bibfnamefont{J.}~\bibnamefont{Bartolom\'e}},
  \bibinfo{author}{\bibfnamefont{J.}~\bibnamefont{Tejada}}, \bibnamefont{and}
  \bibinfo{author}{\bibfnamefont{R.}~\bibnamefont{Ziolo}},
  \bibinfo{journal}{Europhys. Lett.} \textbf{\bibinfo{volume}{35}},
  \bibinfo{pages}{301} (\bibinfo{year}{1996}).

\bibitem[{\citenamefont{Thomas et~al.}(1996)\citenamefont{Thomas, Lionti,
  Ballou, Gatteschi, Sessoli, and Barbara}}]{thoetal96nat}
\bibinfo{author}{\bibfnamefont{L.}~\bibnamefont{Thomas}},
  \bibinfo{author}{\bibfnamefont{F.}~\bibnamefont{Lionti}},
  \bibinfo{author}{\bibfnamefont{R.}~\bibnamefont{Ballou}},
  \bibinfo{author}{\bibfnamefont{D.}~\bibnamefont{Gatteschi}},
  \bibinfo{author}{\bibfnamefont{R.}~\bibnamefont{Sessoli}}, \bibnamefont{and}
  \bibinfo{author}{\bibfnamefont{B.}~\bibnamefont{Barbara}},
  \bibinfo{journal}{Nature} \textbf{\bibinfo{volume}{383}},
  \bibinfo{pages}{145} (\bibinfo{year}{1996}).

\bibitem[{\citenamefont{Maci\`a et~al.}(2009)\citenamefont{Maci\`a, Hernandez,
  Tejada, Datta, Hill, Lampropoulos, and Christou}}]{masiaetal09prb}
\bibinfo{author}{\bibfnamefont{F.}~\bibnamefont{Maci\`a}},
  \bibinfo{author}{\bibfnamefont{J.~M.} \bibnamefont{Hernandez}},
  \bibinfo{author}{\bibfnamefont{J.}~\bibnamefont{Tejada}},
  \bibinfo{author}{\bibfnamefont{S.}~\bibnamefont{Datta}},
  \bibinfo{author}{\bibfnamefont{S.}~\bibnamefont{Hill}},
  \bibinfo{author}{\bibfnamefont{C.}~\bibnamefont{Lampropoulos}},
  \bibnamefont{and} \bibinfo{author}{\bibfnamefont{G.}~\bibnamefont{Christou}},
  \bibinfo{journal}{Phys. Rev. B} \textbf{\bibinfo{volume}{79}},
  \bibinfo{pages}{092403} (\bibinfo{year}{2009}).

\bibitem[{\citenamefont{Chudnovsky and Garanin}(1997)}]{chugar97prl}
\bibinfo{author}{\bibfnamefont{E.~M.} \bibnamefont{Chudnovsky}}
  \bibnamefont{and} \bibinfo{author}{\bibfnamefont{D.~A.}
  \bibnamefont{Garanin}}, \bibinfo{journal}{Phys. Rev. Lett.}
  \textbf{\bibinfo{volume}{79}}, \bibinfo{pages}{4469} (\bibinfo{year}{1997}).

\bibitem[{\citenamefont{Garanin and Chudnovsky}(1997)}]{garchu97prb}
\bibinfo{author}{\bibfnamefont{D.~A.} \bibnamefont{Garanin}} \bibnamefont{and}
  \bibinfo{author}{\bibfnamefont{E.~M.} \bibnamefont{Chudnovsky}},
  \bibinfo{journal}{Phys. Rev. B} \textbf{\bibinfo{volume}{56}},
  \bibinfo{pages}{11102} (\bibinfo{year}{1997}).

\bibitem[{\citenamefont{Luis et~al.}(1997)\citenamefont{Luis, Bartolom\'e,
  Fern\'andez, Tejada, Hern\'andez, Zhang, and Ziolo}}]{luisetal97prb}
\bibinfo{author}{\bibfnamefont{F.}~\bibnamefont{Luis}},
  \bibinfo{author}{\bibfnamefont{J.}~\bibnamefont{Bartolom\'e}},
  \bibinfo{author}{\bibfnamefont{J.~F.} \bibnamefont{Fern\'andez}},
  \bibinfo{author}{\bibfnamefont{J.}~\bibnamefont{Tejada}},
  \bibinfo{author}{\bibfnamefont{J.~M.} \bibnamefont{Hern\'andez}},
  \bibinfo{author}{\bibfnamefont{X.~X.} \bibnamefont{Zhang}}, \bibnamefont{and}
  \bibinfo{author}{\bibfnamefont{R.}~\bibnamefont{Ziolo}},
  \bibinfo{journal}{Phys. Rev. B} \textbf{\bibinfo{volume}{55}},
  \bibinfo{pages}{11448} (\bibinfo{year}{1997}).

\bibitem[{\citenamefont{Garanin and Chudnovsky}(2009)}]{garchu09prl}
\bibinfo{author}{\bibfnamefont{D.~A.} \bibnamefont{Garanin}} \bibnamefont{and}
  \bibinfo{author}{\bibfnamefont{E.~M.} \bibnamefont{Chudnovsky}},
  \bibinfo{journal}{Phys. Rev. Lett.} \textbf{\bibinfo{volume}{102}},
  \bibinfo{pages}{097206} (\bibinfo{year}{2009}).

\bibitem[{\citenamefont{Garanin}(2009)}]{gar09prb}
\bibinfo{author}{\bibfnamefont{D.~A.} \bibnamefont{Garanin}},
  \bibinfo{journal}{Phys. Rev. B} \textbf{\bibinfo{volume}{80}},
  \bibinfo{eid}{014406} (\bibinfo{year}{2009}).

\bibitem[{\citenamefont{Garanin and Chudnovsky}(2008)}]{garchu08prb}
\bibinfo{author}{\bibfnamefont{D.~A.} \bibnamefont{Garanin}} \bibnamefont{and}
  \bibinfo{author}{\bibfnamefont{E.~M.} \bibnamefont{Chudnovsky}},
  \bibinfo{journal}{Phys. Rev. B} \textbf{\bibinfo{volume}{78}},
  \bibinfo{pages}{174425} (\bibinfo{year}{2008}).

\bibitem[{\citenamefont{McHugh et~al.}(2009{\natexlab{c}})\citenamefont{McHugh,
  Jaafar, Sarachik, Myasoedov, Shtrikman, Zeldov, Bagai, and
  Christou}}]{mchughetal09prb}
\bibinfo{author}{\bibfnamefont{S.}~\bibnamefont{McHugh}},
  \bibinfo{author}{\bibfnamefont{R.}~\bibnamefont{Jaafar}},
  \bibinfo{author}{\bibfnamefont{M.~P.} \bibnamefont{Sarachik}},
  \bibinfo{author}{\bibfnamefont{Y.}~\bibnamefont{Myasoedov}},
  \bibinfo{author}{\bibfnamefont{H.}~\bibnamefont{Shtrikman}},
  \bibinfo{author}{\bibfnamefont{E.}~\bibnamefont{Zeldov}},
  \bibinfo{author}{\bibfnamefont{R.}~\bibnamefont{Bagai}}, \bibnamefont{and}
  \bibinfo{author}{\bibfnamefont{G.}~\bibnamefont{Christou}},
  \bibinfo{journal}{Phys. Rev. B} \textbf{\bibinfo{volume}{79}},
  \bibinfo{pages}{052404} (\bibinfo{year}{2009}{\natexlab{c}}).

\bibitem[{\citenamefont{Garanin and Jaafar}(2010)}]{garjaa10prbrc}
\bibinfo{author}{\bibfnamefont{D.~A.} \bibnamefont{Garanin}} \bibnamefont{and}
  \bibinfo{author}{\bibfnamefont{R.}~\bibnamefont{Jaafar}},
  \bibinfo{journal}{Phys. Rev. B} \textbf{\bibinfo{volume}{81}},
  \bibinfo{pages}{180401} (\bibinfo{year}{2010}).

\bibitem[{\citenamefont{Garanin}(2011)}]{gar11acp}
\bibinfo{author}{\bibfnamefont{D.~A.} \bibnamefont{Garanin}}, in
  \emph{\bibinfo{booktitle}{Advances in Chemical Physics}}, edited by
  \bibinfo{editor}{\bibfnamefont{S.~A.} \bibnamefont{Rice}} \bibnamefont{and}
  \bibinfo{editor}{\bibfnamefont{A.~R.} \bibnamefont{Dinner}}
  (\bibinfo{publisher}{John Wiley \& Sons, Inc.}, \bibinfo{address}{Hoboken,
  NJ, USA}, \bibinfo{year}{2011}), vol. \bibinfo{volume}{147}.

\bibitem[{\citenamefont{Chudnovsky}(2004)}]{chu04prl}
\bibinfo{author}{\bibfnamefont{E.~M.} \bibnamefont{Chudnovsky}},
  \bibinfo{journal}{Phys. Rev. Lett.} \textbf{\bibinfo{volume}{92}},
  \bibinfo{pages}{120405} (\bibinfo{year}{2004}).

\bibitem[{\citenamefont{Chudnovsky et~al.}(2005)\citenamefont{Chudnovsky,
  Garanin, and Schilling}}]{chugarsch05prb}
\bibinfo{author}{\bibfnamefont{E.~M.} \bibnamefont{Chudnovsky}},
  \bibinfo{author}{\bibfnamefont{D.~A.} \bibnamefont{Garanin}},
  \bibnamefont{and}
  \bibinfo{author}{\bibfnamefont{R.}~\bibnamefont{Schilling}},
  \bibinfo{journal}{Phys. Rev. B} \textbf{\bibinfo{volume}{72}},
  \bibinfo{pages}{094426} (\bibinfo{year}{2005}).

\bibitem[{\citenamefont{Bokacheva et~al.}(2000)\citenamefont{Bokacheva, Kent,
  and Walters}}]{bokkenwal00prl}
\bibinfo{author}{\bibfnamefont{L.}~\bibnamefont{Bokacheva}},
  \bibinfo{author}{\bibfnamefont{A.~D.} \bibnamefont{Kent}}, \bibnamefont{and}
  \bibinfo{author}{\bibfnamefont{M.~A.} \bibnamefont{Walters}},
  \bibinfo{journal}{Phys. Rev. Lett.} \textbf{\bibinfo{volume}{85}},
  \bibinfo{pages}{4803} (\bibinfo{year}{2000}).

\bibitem[{\citenamefont{Wernsdorfer et~al.}(2006)\citenamefont{Wernsdorfer,
  Murugesu, and Christou}}]{wermugchr06prl}
\bibinfo{author}{\bibfnamefont{W.}~\bibnamefont{Wernsdorfer}},
  \bibinfo{author}{\bibfnamefont{M.}~\bibnamefont{Murugesu}}, \bibnamefont{and}
  \bibinfo{author}{\bibfnamefont{G.}~\bibnamefont{Christou}},
  \bibinfo{journal}{Phys. Rev. Lett.} \textbf{\bibinfo{volume}{96}},
  \bibinfo{pages}{057208} (\bibinfo{year}{2006}).

\bibitem[{\citenamefont{Gomes et~al.}(1998)\citenamefont{Gomes, Novak, Sessoli,
  Caneschi, and Gatteschi}}]{gometal98prb}
\bibinfo{author}{\bibfnamefont{A.~M.} \bibnamefont{Gomes}},
  \bibinfo{author}{\bibfnamefont{M.~A.} \bibnamefont{Novak}},
  \bibinfo{author}{\bibfnamefont{R.}~\bibnamefont{Sessoli}},
  \bibinfo{author}{\bibfnamefont{A.}~\bibnamefont{Caneschi}}, \bibnamefont{and}
  \bibinfo{author}{\bibfnamefont{D.}~\bibnamefont{Gatteschi}},
  \bibinfo{journal}{Phys. Rev. B} \textbf{\bibinfo{volume}{57}},
  \bibinfo{pages}{5021} (\bibinfo{year}{1998}).

\bibitem[{\citenamefont{Garanin}(2010)}]{gar10prbrc}
\bibinfo{author}{\bibfnamefont{D.~A.} \bibnamefont{Garanin}},
  \bibinfo{journal}{Phys. Rev. B} \textbf{\bibinfo{volume}{81}},
  \bibinfo{pages}{220408} (\bibinfo{year}{2010}).

\bibitem[{\citenamefont{Garanin}(2008{\natexlab{a}})}]{gar08prbrc}
\bibinfo{author}{\bibfnamefont{D.~A.} \bibnamefont{Garanin}},
  \bibinfo{journal}{Phys. Rev. B} \textbf{\bibinfo{volume}{78}},
  \bibinfo{pages}{020405(R)} (\bibinfo{year}{2008}{\natexlab{a}}).

\bibitem[{\citenamefont{del Barco et~al.}(2003)\citenamefont{del Barco, Kent,
  Rumberger, Hendrickson, and Cristou}}]{barkenrumhencri03prl}
\bibinfo{author}{\bibfnamefont{E.}~\bibnamefont{del Barco}},
  \bibinfo{author}{\bibfnamefont{A.~D.} \bibnamefont{Kent}},
  \bibinfo{author}{\bibfnamefont{E.~M.} \bibnamefont{Rumberger}},
  \bibinfo{author}{\bibfnamefont{D.~N.} \bibnamefont{Hendrickson}},
  \bibnamefont{and} \bibinfo{author}{\bibfnamefont{G.}~\bibnamefont{Cristou}},
  \bibinfo{journal}{Phys. Rev. Lett.} \textbf{\bibinfo{volume}{91}},
  \bibinfo{pages}{047203} (\bibinfo{year}{2003}).

\bibitem[{\citenamefont{Calero et~al.}(2006)\citenamefont{Calero, Chudnovsky,
  and Garanin}}]{calchugar06prb}
\bibinfo{author}{\bibfnamefont{C.}~\bibnamefont{Calero}},
  \bibinfo{author}{\bibfnamefont{E.~M.} \bibnamefont{Chudnovsky}},
  \bibnamefont{and} \bibinfo{author}{\bibfnamefont{D.~A.}
  \bibnamefont{Garanin}}, \bibinfo{journal}{Phys. Rev. B}
  \textbf{\bibinfo{volume}{74}}, \bibinfo{pages}{094428}
  (\bibinfo{year}{2006}).

\bibitem[{\citenamefont{Leuenberger and Loss}(1999)}]{leulos99epl}
\bibinfo{author}{\bibfnamefont{M.~N.} \bibnamefont{Leuenberger}}
  \bibnamefont{and} \bibinfo{author}{\bibfnamefont{D.}~\bibnamefont{Loss}},
  \bibinfo{journal}{Europhys. Lett.} \textbf{\bibinfo{volume}{46}},
  \bibinfo{pages}{692} (\bibinfo{year}{1999}).

\bibitem[{\citenamefont{Dicke}(1954)}]{dic54}
\bibinfo{author}{\bibfnamefont{R.}~\bibnamefont{Dicke}},
  \bibinfo{journal}{Phys. Rev.} \textbf{\bibinfo{volume}{93}},
  \bibinfo{pages}{99} (\bibinfo{year}{1954}).

\bibitem[{\citenamefont{Chudnovsky and Garanin}(2002)}]{chugar02prl}
\bibinfo{author}{\bibfnamefont{E.~M.} \bibnamefont{Chudnovsky}}
  \bibnamefont{and} \bibinfo{author}{\bibfnamefont{D.~A.}
  \bibnamefont{Garanin}}, \bibinfo{journal}{Phys. Rev. Lett.}
  \textbf{\bibinfo{volume}{89}}, \bibinfo{pages}{157201}
  (\bibinfo{year}{2002}).

\bibitem[{\citenamefont{Chudnovsky and Garanin}(2004)}]{chugar04prl}
\bibinfo{author}{\bibfnamefont{E.~M.} \bibnamefont{Chudnovsky}}
  \bibnamefont{and} \bibinfo{author}{\bibfnamefont{D.~A.}
  \bibnamefont{Garanin}}, \bibinfo{journal}{Phys. Rev. Lett.}
  \textbf{\bibinfo{volume}{93}}, \bibinfo{pages}{257205}
  (\bibinfo{year}{2004}).

\bibitem[{\citenamefont{Abragam and Bleaney}(1970)}]{abrble70}
\bibinfo{author}{\bibfnamefont{A.}~\bibnamefont{Abragam}} \bibnamefont{and}
  \bibinfo{author}{\bibfnamefont{A.}~\bibnamefont{Bleaney}},
  \emph{\bibinfo{title}{Electron {P}aramagnetic {R}esonance of {T}ransition
  {I}ons}} (\bibinfo{publisher}{Clarendon Press}, \bibinfo{address}{Oxford},
  \bibinfo{year}{1970}).

\bibitem[{\citenamefont{Garanin}(2007)}]{gar07prb}
\bibinfo{author}{\bibfnamefont{D.~A.} \bibnamefont{Garanin}},
  \bibinfo{journal}{Phys. Rev. B} \textbf{\bibinfo{volume}{75}},
  \bibinfo{pages}{094409} (\bibinfo{year}{2007}).

\bibitem[{\citenamefont{Garanin}(2008{\natexlab{b}})}]{gar08prb}
\bibinfo{author}{\bibfnamefont{D.~A.} \bibnamefont{Garanin}},
  \bibinfo{journal}{Phys. Rev. B} \textbf{\bibinfo{volume}{77}},
  \bibinfo{pages}{024429} (\bibinfo{year}{2008}{\natexlab{b}}).

\bibitem[{\citenamefont{Carretta et~al.}(2004)\citenamefont{Carretta, Liviotti,
  Magnani, Santini, and Amoretti}}]{caretal04prl}
\bibinfo{author}{\bibfnamefont{S.}~\bibnamefont{Carretta}},
  \bibinfo{author}{\bibfnamefont{E.}~\bibnamefont{Liviotti}},
  \bibinfo{author}{\bibfnamefont{N.}~\bibnamefont{Magnani}},
  \bibinfo{author}{\bibfnamefont{P.}~\bibnamefont{Santini}}, \bibnamefont{and}
  \bibinfo{author}{\bibfnamefont{G.}~\bibnamefont{Amoretti}},
  \bibinfo{journal}{Phys. Rev. Lett.} \textbf{\bibinfo{volume}{92}},
  \bibinfo{pages}{207205} (\bibinfo{year}{2004}).

\bibitem[{\citenamefont{Park et~al.}(2001)\citenamefont{Park, Novotny, Dalal,
  Hill, and Rikvold}}]{parketal01prb}
\bibinfo{author}{\bibfnamefont{K.}~\bibnamefont{Park}},
  \bibinfo{author}{\bibfnamefont{M.~A.} \bibnamefont{Novotny}},
  \bibinfo{author}{\bibfnamefont{N.~S.} \bibnamefont{Dalal}},
  \bibinfo{author}{\bibfnamefont{S.}~\bibnamefont{Hill}}, \bibnamefont{and}
  \bibinfo{author}{\bibfnamefont{P.~A.} \bibnamefont{Rikvold}},
  \bibinfo{journal}{Phys. Rev. B} \textbf{\bibinfo{volume}{65}},
  \bibinfo{pages}{014426} (\bibinfo{year}{2001}).

\bibitem[{\citenamefont{Park et~al.}(2002)\citenamefont{Park, Novotny, Dalal,
  Hill, and Rikvold}}]{parketal02prb}
\bibinfo{author}{\bibfnamefont{K.}~\bibnamefont{Park}},
  \bibinfo{author}{\bibfnamefont{M.~A.} \bibnamefont{Novotny}},
  \bibinfo{author}{\bibfnamefont{N.~S.} \bibnamefont{Dalal}},
  \bibinfo{author}{\bibfnamefont{S.}~\bibnamefont{Hill}}, \bibnamefont{and}
  \bibinfo{author}{\bibfnamefont{P.~A.} \bibnamefont{Rikvold}},
  \bibinfo{journal}{Phys. Rev. B} \textbf{\bibinfo{volume}{66}},
  \bibinfo{pages}{144409} (\bibinfo{year}{2002}).

\end{thebibliography}

\end{document}